\documentclass[lettersize,journal]{IEEEtran}
\usepackage{amsmath,amsfonts}
\usepackage{algorithmic}
\usepackage{algorithm}
\usepackage{array}
\usepackage{textcomp}
\usepackage{stfloats}
\usepackage{url}
\usepackage{verbatim}
\usepackage{graphicx}
\usepackage[numbers]{natbib}
\usepackage{booktabs} 
\usepackage{caption}
\usepackage{multirow}
\usepackage{multicol}
\usepackage{subcaption}
\usepackage[hidelinks]{hyperref}
\usepackage{cleveref}
\usepackage{filecontents}
\usepackage{xcolor}

\newcommand{\prg}[1]{
\noindent
\textbf{#1}.}

\begin{document}

\title{Osmosis Distillation: Model Hijacking with the Fewest Samples}

\author{Yuchen Shi$^{\spadesuit,\dagger}$, Huajie Chen$^{\spadesuit, \clubsuit,\dagger}$, Heng Xu$^\dagger$, Zhiquan Liu$^\ddagger$, Jialiang Shen$^\P$, Chi Liu$^\dagger$, Shuai Zhou$^\dagger$, Tianqing Zhu$^\dagger$, Wanlei Zhou$^\dagger$% <-this % stops a space
\thanks{$^\spadesuit$Equal contribution, $^\clubsuit$Corresponding author. $^\dagger$Faculty of Data Science, City University of Macau.
$^\ddagger$College of Cyber Security, Jinan University.
$^\P$The University of Sydney.
E-mail:\{D23090120503, hjchen, hengxu, shuaizhou, chiliu, tqzhu, wlzhou\}@cityu.edu.mo.
zqliu@vip.qq.com.
shenjial12345@gmail.com.
}
}% <-this % stops a space

\maketitle

\begin{abstract}
 Transfer learning is devised to leverage knowledge from pre-trained models to solve new tasks with limited data and computational resources. 
Meanwhile, dataset distillation has emerged to synthesize a compact dataset that preserves critical information from the original large dataset.
Therefore, a combination of transfer learning and dataset distillation offers promising performance in evaluations.
However, a non-negligible security threat remains undiscovered in transfer learning using synthetic datasets generated by dataset distillation methods, where \textit{an adversary can perform a model hijacking attack with only a few poisoned samples in the synthetic dataset.}
To reveal this threat, we propose \textbf{Osmosis Distillation (OD)} attack, a novel model hijacking strategy that targets deep learning models using the fewest samples. 
The adversary aims to stealthily incorporate a hijacking task into the victim model, forcing it to perform malicious functions without alerting the victim. 
OD attack focuses on efficiency and stealthiness by using the fewest synthetic samples to complete the attack. 
To achieve this, we devise a Transporter that employs a U-Net-based encoder-decoder architecture.
The Transporter generates osmosis samples by optimizing visual and semantic losses to ensure that the hijacking task is difficult to detect. 
The osmosis samples are then distilled into a distilled osmosis set using our specifically designed key patch selection, label reconstruction, and training trajectory matching, ensuring that the distilled osmosis samples retain the properties of the osmosis samples. 
The model trained on the distilled osmosis dataset can perform the original and hijacking tasks seamlessly. 
Comprehensive evaluations on various datasets demonstrate that the OD attack attains high attack success rates in hidden tasks while preserving high model utility in original tasks. 
Furthermore, the distilled osmosis set enables model hijacking across diverse model architectures, allowing model hijacking in transfer learning with considerable attack performance and model utility.
We argue that \textit{awareness of using third-party synthetic datasets in transfer learning must be raised}. 
\end{abstract}


\section{Introduction}
  \begin{figure}[!t]
        \centering
        \includegraphics[width=1.0\linewidth]{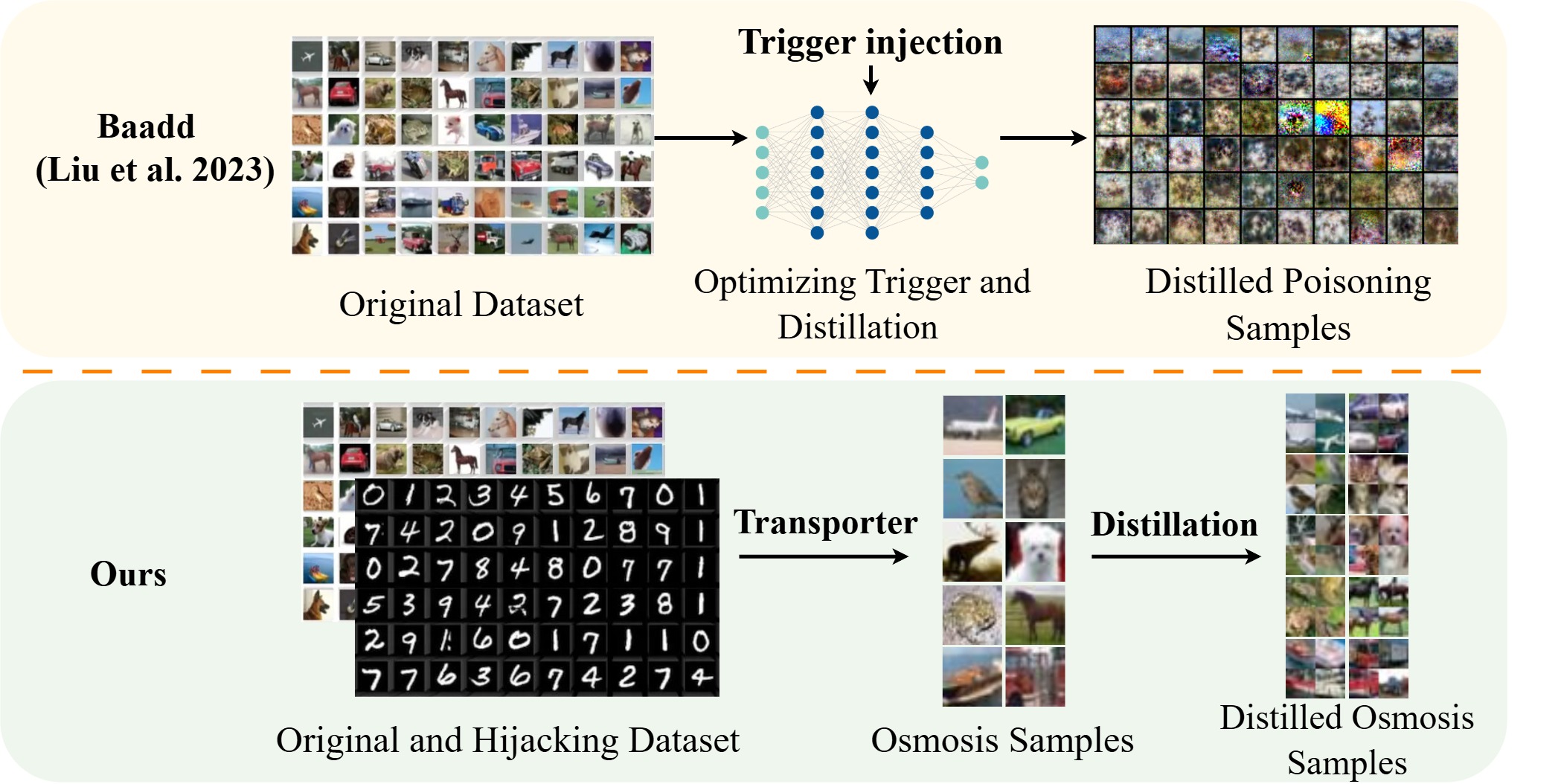}
        \caption{The overview of our work. Different from backdoor attacks, OD attack incorporates a hijacking task into the original task by generating a distilled osmosis dataset that achieves the hijacking task with the fewest samples.}
        \label{fig: overview}
    \end{figure}

\IEEEPARstart{D}{eep} learning relies on large datasets to train models with high predictive accuracy and strong generalization ability. 
However, using large datasets usually poses significant challenges due to high computational costs and a long training time.
To alleviate those issues, methods such as dataset distillation and transfer learning have been proposed.

Dataset distillation is a process that extracts essential information from a large dataset to create a much smaller synthetic dataset. 
This distilled dataset retains the key characteristics of the original, enabling models trained on it to achieve performance comparable to those trained on the full dataset \cite{10273632}. 
Meanwhile, transfer learning allows models to adapt knowledge acquired from a source domain to a different target domain by leveraging shared latent structures such as features.
    
To enhance training efficiency and mitigate computational resource consumption, users are opting to use third-party distilled datasets for fine-tuning pre-trained models obtained from open-source repositories.
However, this introduces novel security and privacy vulnerabilities in the real world, particularly when faced with model hijacking attacks.
These attacks make the model that is fine-tuned on the synthetic dataset execute hijacking tasks unwittingly while preserving performance on the original tasks. 
Notably, the hijacking task defined by the adversary may involve serious illegal activities. 
Consequently, model hijacking poses risks of parasitic computation and stealthy criminality \cite{salem2022get}. 
Moreover, existing model hijacking attacks still require many hijacking samples to compromise the victim model, and there is relatively little exploration of distilled datasets in mounting such attacks.
\prg{Our Work} To reveal this undiscovered threat of using dataset distillation in transfer learning, we aim to combine model hijacking and dataset distillation to enable such attacks with a minimal number of hijacking samples and explore the feasibility of achieving them via distilled datasets. As shown in \autoref{fig: overview}, unlike typical backdoor attacks, our method does not need triggers or intend to induce misclassifications in machine learning models.
Instead, it aims to force the model to execute the hijacking task specified by the adversary.
The proposed attack method comprises two important steps: \textbf{O}smosis and \textbf{D}istillation.
Therefore, we call this method \textbf{OD attack}.

In OD attack, we design a model named Transporter that is built upon an encoder-decoder architecture. 
The Transporter is used to disguise osmosis samples as benign samples. 
To ensure that osmosis samples are visually similar to benign samples in the original dataset and semantically similar to osmosis samples in the hijacking dataset, the Transporter is trained using two loss functions: visual loss and semantic loss. 
The visual loss ensures that the osmosis samples visually resemble the benign samples, while the semantic loss ensures that they maintain semantic similarity to the hijacking samples. 
After the generation of the osmosis samples, the distillation stage starts.
Initially, each osmosis sample is cropped into multiple patches of equal size.
We then compute a realism score for each of the patches, and select the patch with the highest score as the key patch. These key patches are subsequently used to reconstruct a complete synthetic image. 
Following this, we perform label reconstruction, employing soft labels and training trajectory matching to guarantee that the distilled osmosis samples maintain the characteristics of the hijacking samples.
The target model that is trained on such a distilled osmosis dataset (DOD) eventually possesses the ability to perform both the original task and the hijacking task specified by the adversary with high accuracy.

Our contributions are summarized as follows:
    \begin{itemize}
        \item To the best of our knowledge, our work is the first to reveal potential risks in transfer learning using synthetic datasets generated by dataset distillation.

        \item Our proposed OD attack uses distilled osmosis samples for the hijacking task, ensuring that the adversary can use the fewest samples to launch model hijacking attacks.
        This approach also ensures that the synthetic samples are difficult to detect, which guarantees the stealthiness of the attack.

        \item Experimental results indicate that a distilled osmosis dataset with only fifty samples in each class can effectively ensure the attack success rate and model utility of model hijacking attacks.
    \end{itemize}

\section{Preliminaries and Related Work}

\begin{table}[!t]
    \centering
    \caption{Notations}
    \label{tab: notations}
    \begin{tabular}{c|l}
        \toprule Symbols      & Definitions                             \\
        \midrule 
        $x_{o}$               & Original samples                        \\
        $x_{h}$               & Hijacking samples                       \\
        $x_{c}$               & Osmosis samples                         \\
        $x_{c\_syn}$          & Distilled osmosis samples               \\
        $\mathcal{D}_{o}$     & Original dataset                        \\
        $\mathcal{D}_{h}$     & Hijacking dataset                       \\
        $\mathcal{D}_{c\_syn}$     & Distilled osmosis dataset          \\
        $\mathcal{F}_{(x_c)}$ & Feature extractor for osmosis samples   \\
        $\mathcal{F}_{(x_h)}$ & Feature extractor for hijacking samples \\
        \bottomrule
    \end{tabular}
\end{table}
All notations used in this work are listed in \autoref{tab: notations}.

\subsection{Transfer learning}
The primary objective of transfer learning is to leverage tasks and knowledge from the source domain ($\mathcal{D}_{S}= \{( x_{S1},y_{S1}), \ldots ,(x_{S_{n_{S}}}, y_{S_{n_{S}}}) \}$)  to improve the target predictive function ($f_{T}(\cdot)$) in the target domain ($\mathcal{D}_{T}= \{( x_{T1},y_{T1}), \ldots ,(x_{T_{n_{T}}}, y_{T_{n_{T}}}) \}$). 
Transfer learning is often used in cases where the source domain and the target domain feature spaces or marginal distributions are different.
Transfer learning uses source tasks that contain abundant data to obtain features and knowledge, so as to reduce the amount of labeled data required for the target task to improve training efficiency and model performance  \cite{lu2015transfer,5288526,shen2025dino}.

\subsection{Dataset Distillation}
Dataset distillation aims to compress a large-scale dataset ($\mathcal{D}_{real}$) into a smaller synthetic dataset ($\mathcal{D}_{syn}$) \cite{wang2018dataset}. 
Its objective can be formulated as:
    \begin{equation}
    \label{eq: distillation}
    \mathcal{D}^*_{syn}= \arg\min_{\mathcal{D}_{syn}} \mathcal{L}(\mathcal{D}_{syn},\mathcal{D}_{real}).
    \end{equation}
    
To improve efficiency, Zhao et al. \cite{zhao2020dataset, zhao2023dataset} introduced the first-order gradient matching and later distribution matching techniques. Cazenavette et al. \cite{cazenavette2022dataset} proposed the trajectory matching technique, aligning optimization paths between real and synthetic data. 
Other approaches include patch-based image and soft label reconstruction \cite{sun2024diversity}, neural tangent kernel regression \cite{nguyen2020dataset, nguyen2021dataset}, and final-layer regression \cite{zhou2022dataset}.

\subsection{Backdoor Attack}
In backdoor attacks, the adversary manipulates the training process of the victim model to implant backdoors.
Most commonly, the adversary designs a trigger and injects it into the training data, causing the model to predict a specified label upon encountering inputs containing the trigger. 
Gu et al. \cite{gu2017badnets} first proposed BadNets, a method to backdoor machine learning models using a blank pixel as a trigger to misclassify backdoor inputs as target labels. 
Salem et al. \cite{salem2022dynamic} proposed using dynamic trigger to execute backdoor attacks. 
Further, various backdoor attack methods have been proposed for dataset distillation \cite{liu2023backdoor}, diffusion models \cite{chou2023backdoor, chen2023trojdiff}, image classification \cite{doan2021backdoor, doan2021lira}, natural language processing models \cite{schuster2021you}, transfer learning \cite{yao2019latent}, and others \cite{saha2020hidden,li2021invisible,rakin2020tbt,wang2020attack,zhao2020clean,chen2025queen}.
    
\subsection{Model Hijacking Attacks}

Salem et al. first proposed model hijacking attacks \cite{salem2022get} as a training-time attack strategy. 
The goal of model hijacking is to covertly redirect the functionality of a victim model from its intended task to an adversary-specified task, while preserving the victim model’s performance on the original task to avoid detection.
The model hijacking method proposed by Salem et al.    
utilizes a Camouflager based on an encoder-decoder architecture to embed hijacking samples ($x_h$) into original samples ($x_o$), thereby generating camouflaged samples ($x_c$). 
In this procedure, visual and semantic losses are employed to ensure that camouflaged samples closely resemble original samples in appearance while maintaining similarity to the hijacking samples. The objective is formulated as follows:

\begin{equation}
    \mathcal{L}(x_o, x_h, x_c) = \min  \left( \|x_{c}- x_{o}\| + \|\mathcal{F}
    (x_{c}) - \mathcal{F}(x_{h})\| \right)
\end{equation}

Then, a large number of camouflaged samples form the camouflaged dataset. Together with the original dataset, they constitutes the poisoned dataset. 
The poisoned dataset is used to train the victim model to incorporate the hijacking task into the original task.

Furthermore, \cite{si2023two} extended model hijacking attacks to text generation and  classification models, thereby broadening the scope of such assaults.
Additionally, other relevant work focus on federated learning etc.\cite{chow2023stdlens,zhang2024vera,DBLP:conf/aaai/HeCPLZWLJ25,chen2025stand}.

\section{Problem Formulation}
        \begin{figure*}[!t]
        \centering
        \includegraphics[width=\textwidth]{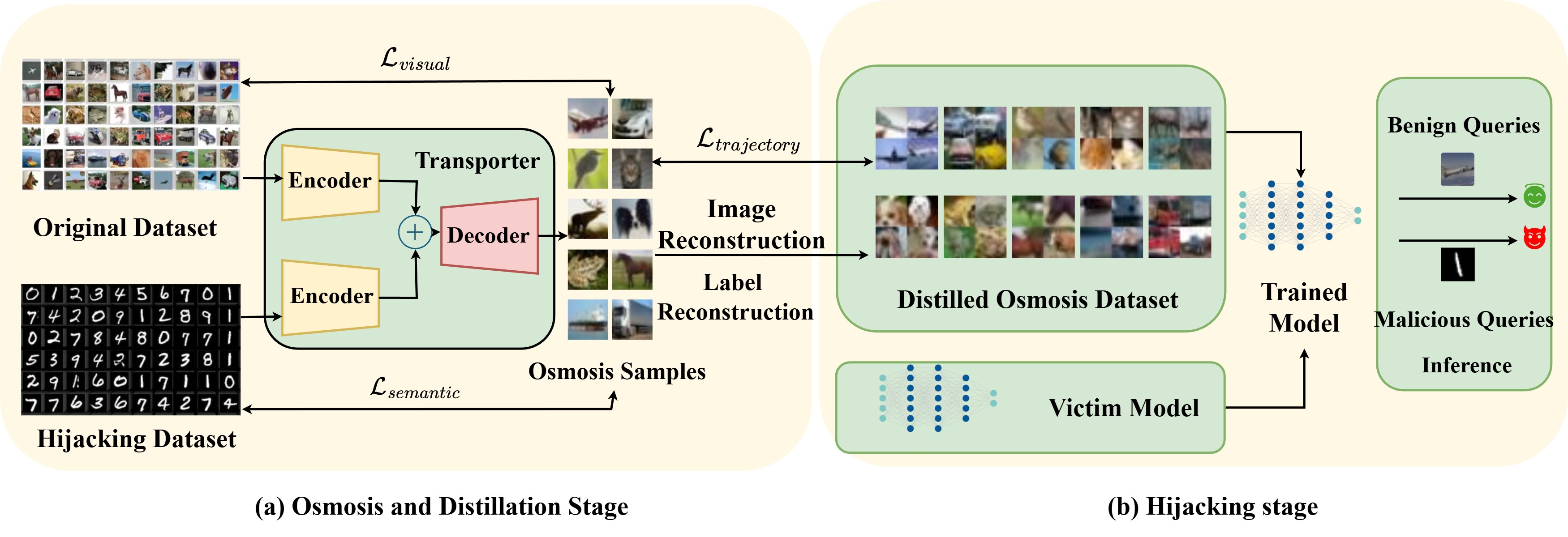}
        \caption{The workflow of OD attack. 
        In stage (a), a Transporter is utilized to embed the hijacking task into the original task, producing osmosis samples, which are then distilled using image reconstruction, label reconstruction and training trajectory matching. 
        In this stage (b), we solely use the distilled osmosis dataset for training the target model. The trained model executes either the original task or hijacking task based on varying queries.}
        \label{fig: workflow}
    \end{figure*}
    
In our study, we consider two parties: an adversary and a victim.
\begin{itemize}
        \item \textbf{Adversary:} An adversary is defined as a malicious entity that actively manipulates the training process, potentially acting as the provider of third-party synthetic datasets generated by some dataset distillation algorithm.
        Its objectives encompass exploiting the victim's computational resources to execute proprietary tasks and imposing legal or ethical risks on the victim through the enforcement of illicit activities.

     \item \textbf{Victim:} A victim could be an individual model owner or a company that wants to use synthetic datasets to speed up model fine-tuning. 
     They are likely to choose third-party distilled datasets from open-source platforms.
     As the victim model has high performance on the original task, the victim is less likely to notice the hijacking task.
     Consequently, the victim faces the risk of delivering unauthorized services and parasitic computation.
\end{itemize}

\subsection{Threat Model}
\label{threat model}
\prg{Adversary's Goals} 
The goal of the adversary is to incorporate a hijacking task defined by the adversary into a victim model.
The victim model preserves its utility with regard to its original task, while having considerable performance on the hijacking task.
The victim must not notice the existence of the hijacking task.
To this end, OD attack must have the following four properties:
\textbf{P1: Effectiveness.} Effectiveness requires the victim model to have high performance in both the original task and the hijacking task. 
Furthermore, the existence of the hijacking task should not affect the performance of the original task.
\textbf{P2: Efficiency.} It is expected that OD attack will be effective with the fewest samples to accelerate the fine-tuning process.
\textbf{P3: Stealthiness.} Distilled osmosis samples are expected to exhibit a high degree of visual similarity to original samples.
\textbf{P4: Transferability.} Distilled osmosis datasets should support transfer learning regardless of model architectures or optimization algorithms.

\prg{Adversary's Knowledge} 
The adversary's knowledge is strictly limited. The victim model's architecture, training algorithm and parameters are unknown to the adversary.
Nonetheless, the adversary has knowledge about all existing public datasets, dataset distillation algorithms, and platforms for dataset providers.

\prg{Adversary's Capability}
The adversary cannot interfere with the training process of the victim model.
However, the adversary can control the original dataset and the hijacking dataset that are used to generate the DOD.
All accessible online datasets are available to the adversary for collection.
The adversary is also allowed to produce private datasets.
Furthermore, the adversary can determine the algorithm that is used to generate the DOD.
The adversary can select online platforms to release the DOD.

\prg{Victim's Goal} 
The victim's goal is to swiftly train a model and execute the original task precisely.

\prg{Victim's Knowledge} 
The victim has knowledge about all existing model architectures, training algorithms, and publicly accessible datasets.
The victim does not know whether the dataset that is used to fine-tune the victim model contains harmful contents or not.

\prg{Victim's Capability} 
The victim can select any model architectures, training algorithms to train the victim model.
The third-party dataset for training the victim model is assumed to be the DOD generated by the adversary, but the victim can still locally manipulate the DOD.

\section{OD}
As depicted in \autoref{fig: workflow}, OD attack comprises two primary components.
The first is the osmosis and distillation stage, which aims to generate distilled osmosis samples that visually resemble the samples in the original dataset while semantically aligning with the samples in the hijacking dataset.
The second is the hijacking stage, encompassing model training and inference to execute the OD attack.

\subsection{Osmosis and Distillation Stage}
\label{od}
\subsubsection{Transporter}
    To embed the information of hijacking samples ($x_h$) into original samples ($x_o$), we devise the Transporter based on the encoder-decoder framework grounded in the U-Net architecture.
    In OD attack, the structure comprises two encoders and a single decoder.
    The first encoder process the original samples, while the second handles the hijacking samples.
    Outputs from both encoders are then concatenated to form the decoder's input. 
    The resulting decoder outputs are osmosis samples, which exhibit visual resemblance to the original samples and semantically similar hijacking samples.

    To ensure that the osmosis samples visually resemble the original samples while semantically aligning with the hijacking samples, we design the visual and semantic loss functions in the training stage of the Transporter.

    \prg{Visual loss} The visual loss function computes the L1 distance between the osmosis samples generated by the Transporter and the original samples.
    This loss function serves to ensure that the osmosis samples exhibit a visual resemblance to the original samples. The visual loss is defined as
\begin{equation}
    \label{eq: visual loss}
    \mathcal{L}_{\text{visual}}= \min \left\| x_{c}- x_{o}\right\|.
\end{equation}

    \prg{Semantic loss} The semantic loss operates at the feature level rather than the visual level, a feature extractor is required to capture the characteristics of the hijacking samples. 
    This extractor can be formed by intermediate layers from any classifier model.
    Given our assumption that the adversary lacks access to any information about the victim model, we opt for a pre-trained model as the feature extractor.
    Subsequently, the extracted features of the osmosis samples ($\mathcal{F}_{(x_c)}$) and those of the hijacking samples ($\mathcal{F}_{(x_h)}$) are utilized to compute the L1 distance.
    The semantic loss is defined as
\begin{equation}
    \label{eq: semantic loss}
    \mathcal{L}_{\text{semantic}}= \min\left \| \mathcal{F}_{(x_c)}- \mathcal{F}_{(x_h)}\right\|.
\end{equation}

\subsubsection{Osmosis}
Prior to initiating the Transporter, the adversary must define a mapping function that associates each label in the original dataset with a corresponding label in the hijacking dataset. 
A straightforward approach is to map the $i^{th}$ label from the original dataset to the $i^{th}$ label in the hijacking dataset, without considering the underlying semantic differences between the labels.
It is important to note that the OD attack is independent of the mapping method, the adversary is capable of creating the mapping freely.

After determining the mapping relationship, the adversary can proceed to the osmosis stage. 
At this stage, the visual loss and semantic loss functions previously mentioned are employed to train the Transporter. 
To balance the trade-off between the visual loss and the semantic loss, and thus regulate the interplay between the original task and the hijacking task, we introduce two parameters, denoted as $\lambda_{v}$ and $\lambda_{s}$, which function as weighting coefficients.
The entire loss function for training the Transporter is defined as
\begin{equation}
    \label{eq: total}
    \mathcal{L}(x_{c}, x_{o}, x_{h}) = \lambda_{v}\|x_{c}- x_{o}\| + \lambda_{s}\|\mathcal{F}(x_{c}) - \mathcal{F}(x_{h})\|
\end{equation}

    \begin{algorithm}[!t]
        \caption{OD Attack--Osmosis}
        \label{alg1}
        \textbf{Input:}Original dataset
            $\mathcal{D}_{o}= \{(x_{o}, y_{o})\}$, Hijacking dataset $\mathcal{D}
            _{h}= \{(x_{h}, y_{h})\}$, Label mapping $m: y_{o}\to y_{h}$, 
            transporter $\mathcal{T}$\\
        \textbf{Parameters:}  $\lambda_{v}$, $\lambda_{s}$
            $N$ \\
        \textbf{Output:}  Osmosis samples $x_c$
        \begin{algorithmic}[1]
        \FOR{each epoch}
        \STATE label mapping $y_{h}= m(y_{o})$ 
        \STATE Generate osmosis
            sample $x_{c}= \mathcal{T}(x_{o}, x_{h})$
        \STATE Optimize: $\mathcal{L}= \lambda_{v}\mathcal{L}_{\text{visual}}+ \lambda_{s}\mathcal{L}_{\text{semantic}}$
        \ENDFOR

        \STATE \textbf{return} Osmosis samples $x_c$
    \end{algorithmic}
\end{algorithm}

    \subsubsection{Distill osmosis samples}
    Having obtained the osmosis samples, we proceed to the distillation stage.
    The purpose of this stage is to significantly reduce the number of osmosis samples, and to ensure that the hijacking task remains effective. 
    To guarantee the realism of the osmosis samples after distillation, we first crop each osmosis sample to create patches. 
    Then, we calculate the realism score for each patch using \autoref{eq: realism score} and select the patch with the highest score as the key patch for image synthesis.
    The realism score is defined as
\begin{equation}
    \label{eq: realism score}
    S = -\ell (\phi_{op}(\mathbf{x_c}), \phi_{h}(\mathbf{x_c})) - \ell (\phi_{op}(\mathbf{x_c}),y),
\end{equation}
    where, $\phi_{op}$ is a pre-trained observer model and $\phi_{h}$ is a human observer.
    The human observer is conceptualized as a static mapping grounded in prior knowledge. 
    Given that the original training dataset is annotated by humans, the label $y_i$ corresponding to each samples $x_i$ inherently represents the human observer's judgment of the image.
    Consequently, when osmosis samples are cropped into a patch, provided that the patch retains the core features of the samples, the human observer's classification of the patch remains consistent with the corresponding class.
    This implicitly indicates that the patch possesses features aligned with human cognition.
    Human observers should perceive osmosis samples as original benign samples, given the stealthiness of the attack.

    After obtaining the key patches, we select N key patches for each class and concatenate them into a synthetic image. 
    The synthetic image matches the resolution of the original image. 
    Further, we use soft labels to relabel the synthetic images. 
    The model learns from these reconstructed labels, eventually generating osmosis samples composed of N patches. These samples possess high realism and have reconstructed labels.

    Distillation is a double-edged sword. 
    To ensure that the distilled osmosis samples retain the features of the osmosis samples, a weight trajectory loss is introduced. 
    By minimizing the differences in training trajectories between the distilled osmosis samples and the osmosis samples, this process makes models trained on the DOD produce training weight trajectories that are similar to those of models trained on the set of the osmosis samples.
The weight trajectory loss is defined as
\begin{equation}
    \label{eq: trajectory}
        \mathcal{L}_{\text{trajectory}}(\mathcal{D}_{c\_syn}, \mathcal{D}_{c}) = \frac{\left\| \hat{\theta}_{t+i}-
        \theta_{t+g}^{*}\right\|_{2}^{2}}{\left\| \theta_{t}^{*}- \theta_{t+g}^{*}\right\|_{2}^{2}},
\end{equation}
    where $\theta_{t}^{*}$ is training trajectory of the set of the osmosis samples and $\hat{\theta}_{t}$ is that of the distilled osmosis dataset.

\begin{algorithm}[!t]
    \caption{OD Attack--Distillation}
    \label{alg2}
    \textbf{Input}:Original dataset $\mathcal{D}_{o}= \{(x_{o}, y_{o})\}$, The set of osmosis samples $\mathcal{D}_{c}= \{(x_{h}, y_{h})\}$, Observer models $\phi_{o_p}$, $\phi_{h_p}$ \\
    \textbf{Parameters}: Epoch $N$
    \textbf{Output}: Distilled osmosis samples $o_{c\_syn}$
    \begin{algorithmic}[1]
            \FOR{each class $c$ in hijacking task} 
            \STATE Select osmosis samples $\{(x_{c}, y_{h})\}$ where $y_{h}$ corresponds to class $c$ \FOR{each $x_{c}$}
            \STATE Crop $x_{c}$ into patches $\{p_{i}\}$ \STATE Compute
            $\mathcal{S}= -\ell(\phi_{o_p}(x_{o}), \phi_{h_p}(x_{c})) - \ell(\phi
            _{o_p}(x_{c}), y_{h})$
            \ENDFOR 
            \STATE Select top $N$ patches for class $c$ 
            \STATE Reconstruct image $x_{c}$ by concatenating patches 
            \ENDFOR 
            \STATE Minimize $\mathcal{L}_{\text{trajectory}}(\mathcal{D}_{c\_syn}, \mathcal{D}_{c})$ for distilled samples 
            \STATE \textbf{return} Distilled osmosis samples $o_{c\_syn}$
     \end{algorithmic}
\end{algorithm}

    \subsection{Hijacking Stage}
    After completing the distillation process, the distilled osmosis samples form a compact DOD.
    The DOD is used to fine-tune a pre-trained model. 
    Since the distilled osmosis samples encapsulate information from both the original samples and the hijacking samples, the victim model trained on this distilled dataset can not only perform the original task but also perform the adversary-specified hijacking task.
    Consequently, the hijacking task is covertly integrated into the victim model, transforming it into a victim model.
    When deployed, the victim model can perform the original task well and accurately for benign inputs. However, when exposed to malicious input, the model triggers the adversary-specified hijacking task.

\section{Experiment}
\subsection{Dataset Description}
To show the effectiveness of our schemes, we choose five benchmark datasets for evaluation: 
\begin{itemize}
    \item \textbf{MNIST} \cite{lecun-mnisthandwrittendigit-2010}: MNIST is a dataset of handwritten digits from 0 to 9. Each sample is a gray-scale image of $28 \times 28$ pixels. In our experiments we resize it to $32 \times 32$ pixels.

    \item \textbf{SVHN} \cite{netzer2011reading}: SVHN is a dataset comprising digit images cropped from street view imagery, containing over 600,000 labeled digits. It consists of ten classes, corresponding to the digits 0 to 9, with each sample being a $32 \times 32$ pixels RGB image.

    \item \textbf{CIFAR-10}  \cite{krizhevsky2009learning}: CIFAR-10 is a dataset of images from 10 distinct object classes. Each sample is an RGB image of $32 \times 32$ pixels. 

    \item \textbf{CIFAR-100} \cite{krizhevsky2009learning}: CIFAR-100 is a dataset of images from 100 distinct object classes. Each class contains 600 RGB images of $32 \times 32$ pixels.

    \item \textbf{Tiny-ImageNet} \cite{Le2015TinyIV}: Tiny-ImageNet is a dataset of 200 distinct classes of images. Each class contains 500 training RGB images, each with a resolution of $64 \times 64$ pixels.

    \item \textbf{ImageNet-Subset} : To validate the effectiveness of OD attacks on high-resolution images, we constructed a subset of the ImageNet-1K dataset \cite{5206848}. 
    Specifically, with the image resolution fixed at $224 \times 224$, we randomly selected 200 classes from the original dataset.
    These classes were subsequently partitioned into two disjoint sets, 100 classes allocated to the hijacking task and the remaining 100 classes assigned to the original task.
\end{itemize}

\subsection{Experiment Details}
\subsubsection{Models} The pre-trained MobileNetV2 \cite{Sandler_2018_CVPR} is employed as the feature extractor.
ResNet18 \cite{he2016deep} and VGG16 \cite{SimonyanZ14a} are employed as the architectures of the victim models.

\subsubsection{Baselines}
In this work, we select two SOTA model hijacking attacks targeting image tasks as baselines.
The first baseline is the Chameleon attack \cite{salem2022get}, which was the first to systematically propose a hijacking framework for image classification models.
This research established an attack paradigm based on a Camouflager.
The second is CAMH \cite{DBLP:conf/aaai/HeCPLZWLJ25}, which offers flexibility compared to the Chameleon attack. 
It demonstrates that a small, general-purpose model can be hijacked to process extremely complex classification tasks, thereby expanding the potential attack surface of parasitic computing.

\subsubsection{Evaluation Metrics}
    \begin{itemize}
        \item \textbf{Utility}: The utility of the victim model is its test accuracy on the original test set. 
        The higher the utility, the closer the performance of the victim model is to that of the clean model on the original task. 
        This suggests greater stealthiness of the hijacking task embedded in the OD-attacked distilled dataset.
        Consequently, this increases the likelihood of the distilled dataset being used.

        \item \textbf{Attack Success Rate (ASR)}: The ASR is calculated by its accuracy on the hijacking test set. 
        The higher the ASR, the stronger the attack, underscoring the model's capability to precisely execute the hijacking task as designed by the adversary.
    \end{itemize}

\subsubsection{Implementation}
The Adam optimizer is used for training the victim model.
During training, the label mapping
was defined by random pairing of original and hijacking samples.
We set 100 epochs for training Transporter and 300 epochs for distilling the osmosis samples.
Additionally, the learning rate was set to 0.01, with a batch size of 64. 
In addition, unless otherwise specified, the IPC in our experiments is set to 50 by default and all experimental code is based on the PyTorch framework, and all experiments are conducted on single NVIDIA A100 GPU.

\begin{figure*}[!t]
    \centering
    \includegraphics[width=\linewidth]{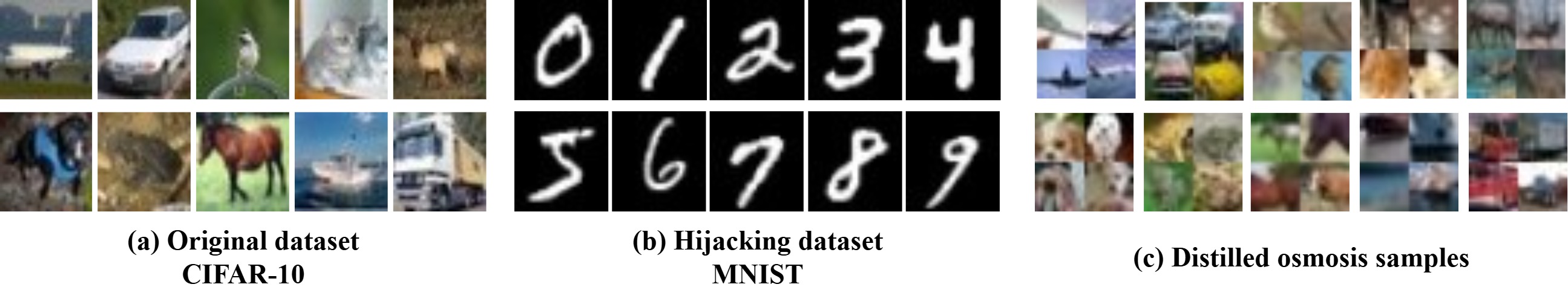}
    \caption{Visualization of the output of OD attack. Figure (a) shows the samples of the Original dataset, figure (b) shows the samples of the hijacking dataset, and figure (c) shows the distilled osmosis samples.}
    \label{fig: od samples visualization}
\end{figure*}

\begin{figure*}[!t]
    \centering
    \begin{subfigure}[b]{0.32\textwidth}
        \centering
        \includegraphics[width=\linewidth]{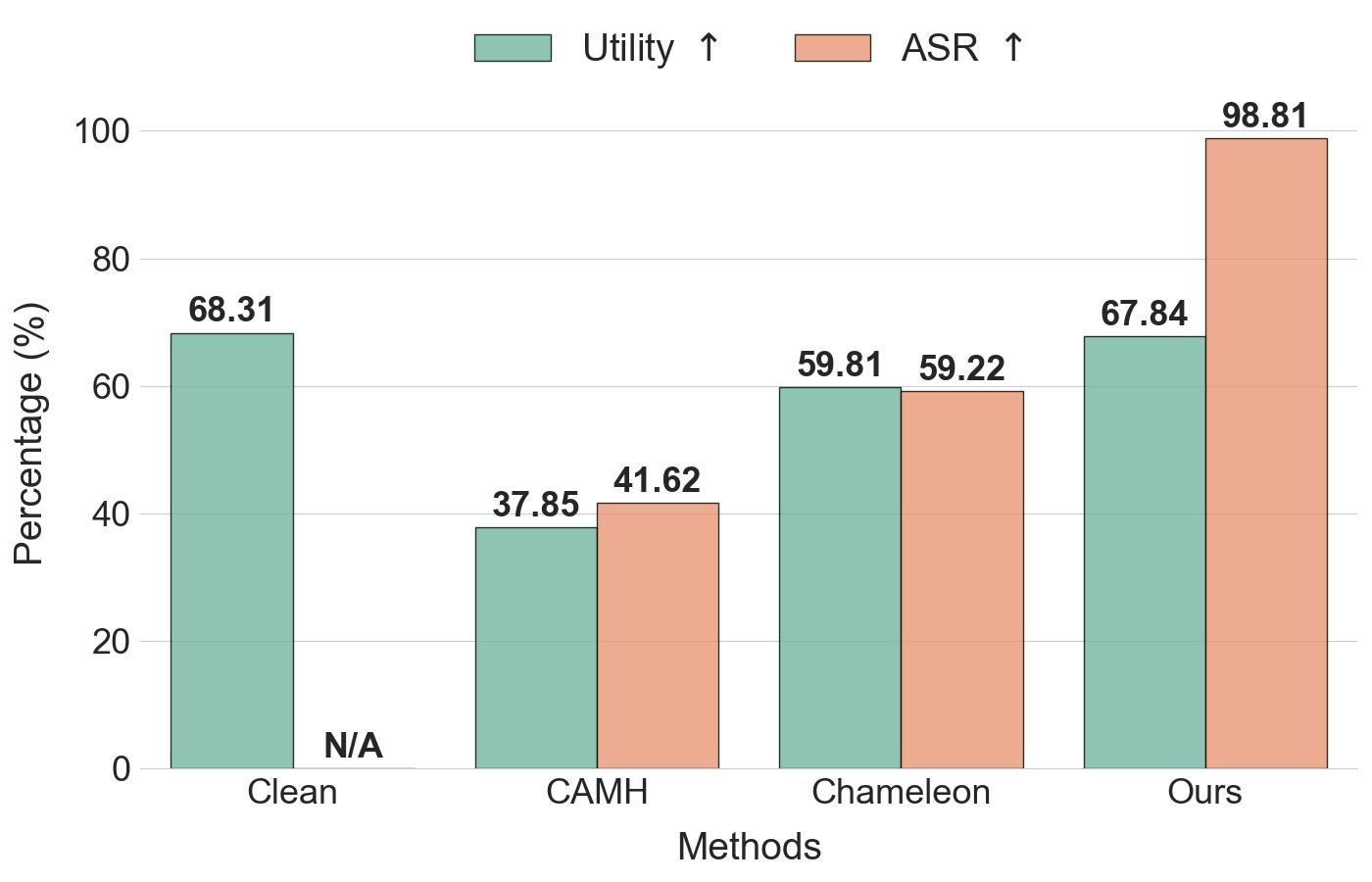}
        \caption{CIFAR-10, MNIST}
        \label{fig:baseline_cifar10_mnist_res18}
    \end{subfigure}
    \hfill
    \begin{subfigure}[b]{0.32\textwidth}
        \centering
        \includegraphics[width=\linewidth]{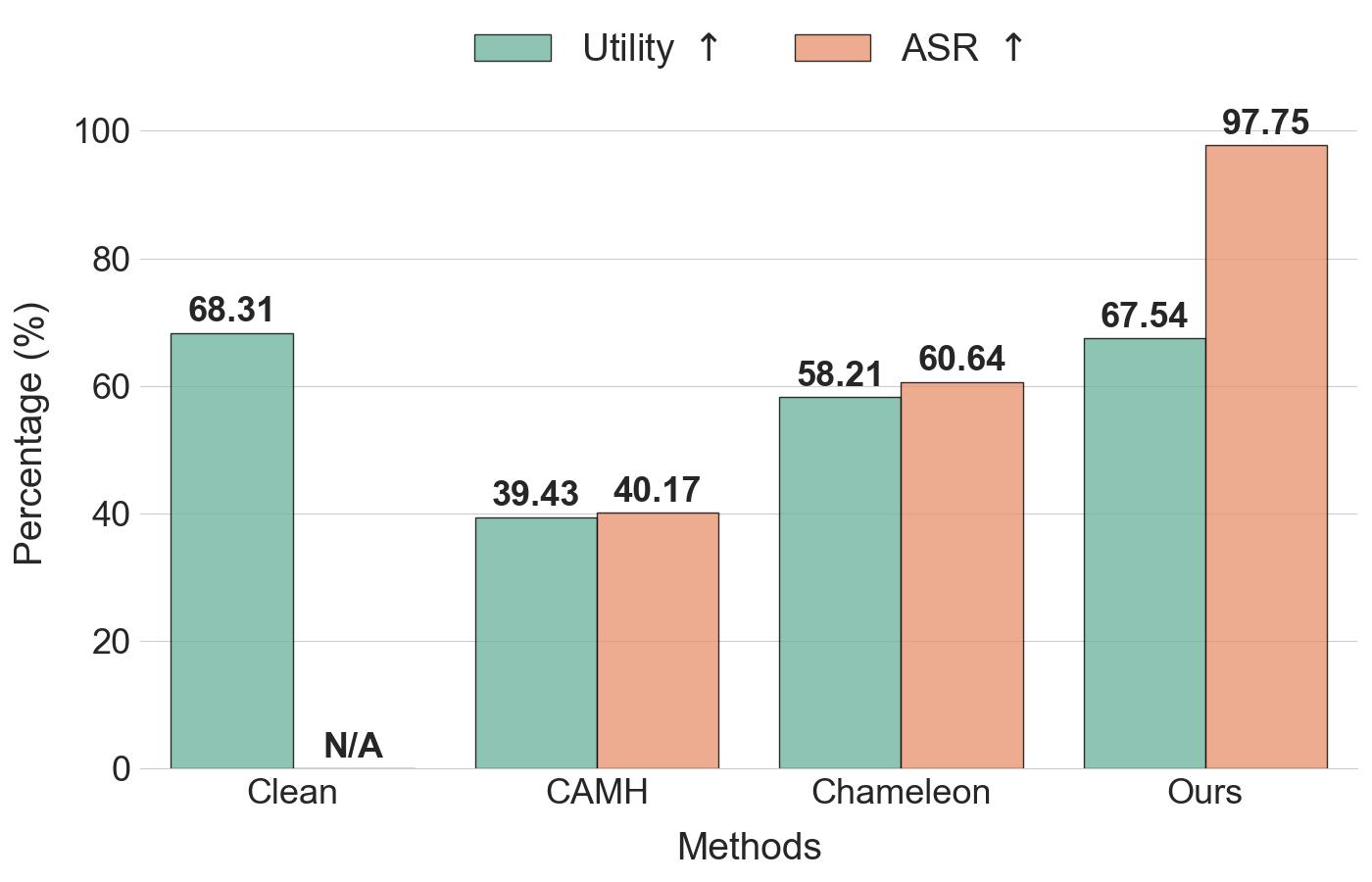}
        \caption{CIFAT-10, SVHN}
        \label{fig:baseline_cifar10_svhn_res18}
    \end{subfigure}
    \hfill
    \begin{subfigure}[b]{0.32\textwidth}
        \centering
        \includegraphics[width=\linewidth]{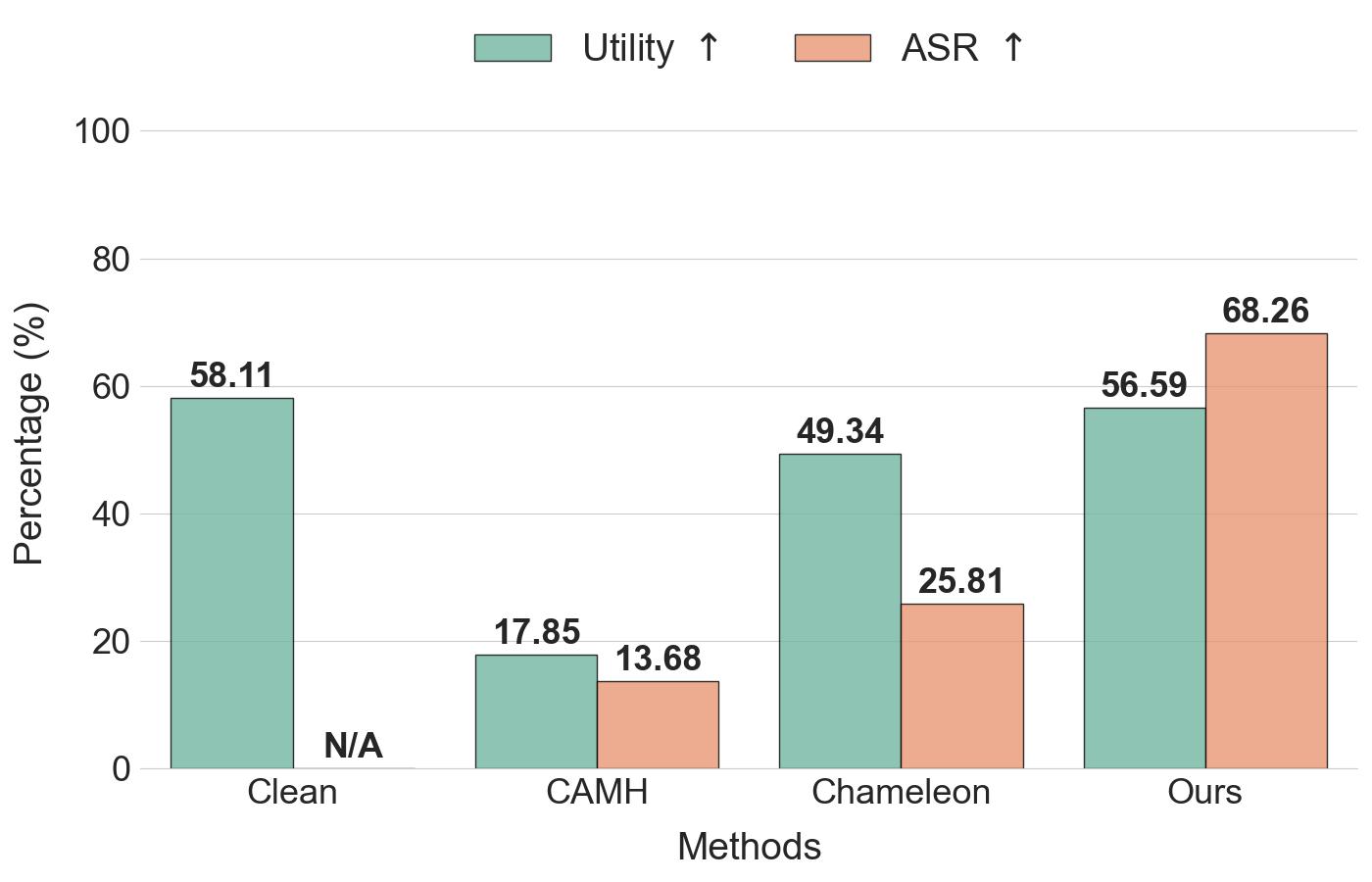}
        \caption{CIFAR-100, Tiny-ImageNet}
        \label{fig:cham_cifar100_tiny_res18}
    \end{subfigure}
    
    \par\medskip 
    \begin{subfigure}[b]{0.32\textwidth}
        \centering
        \includegraphics[width=\linewidth]{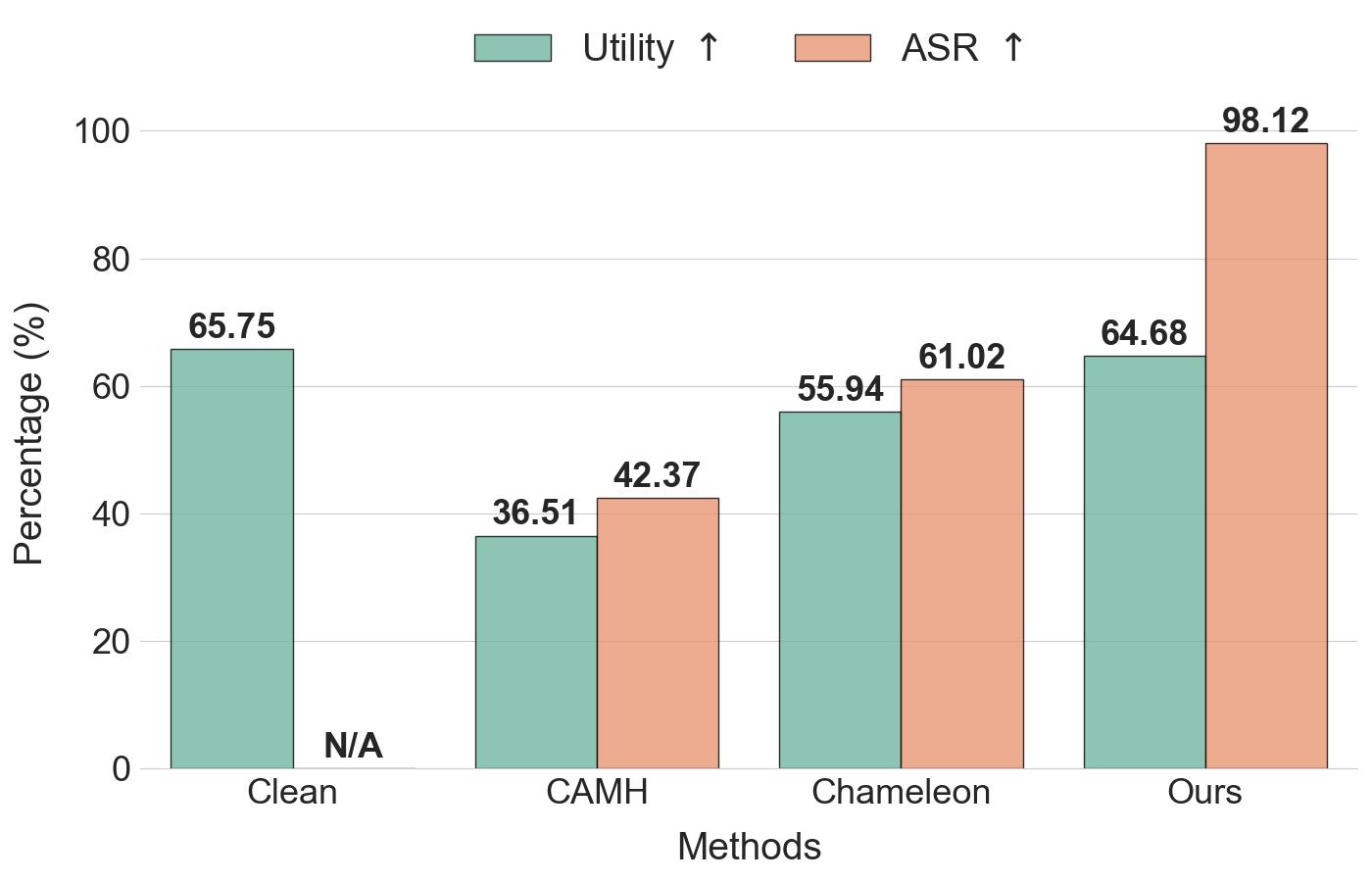}
         \caption{CIFAR-10, MNIST}
         \label{fig:cham_cifar10_mnist_vgg}
    \end{subfigure}
    \hfill
    \begin{subfigure}[b]{0.32\textwidth}
        \centering
        \includegraphics[width=\linewidth]{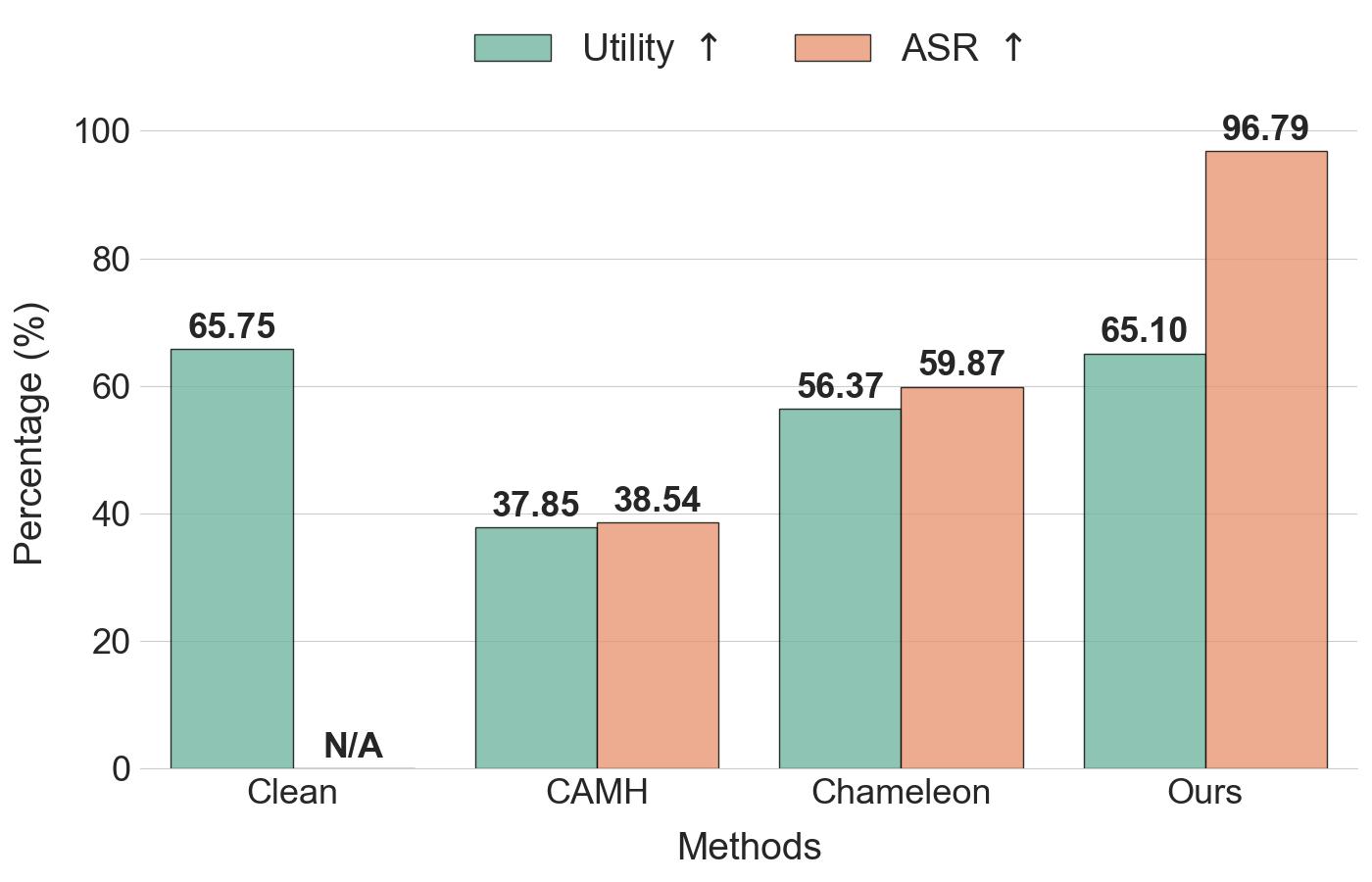}
        \caption{CIFAT-10, SVHN}
        \label{fig:cham_cifar10_svhn_vgg}
    \end{subfigure}
    \hfill
    \begin{subfigure}[b]{0.32\textwidth}
        \centering
        \includegraphics[width=\linewidth]{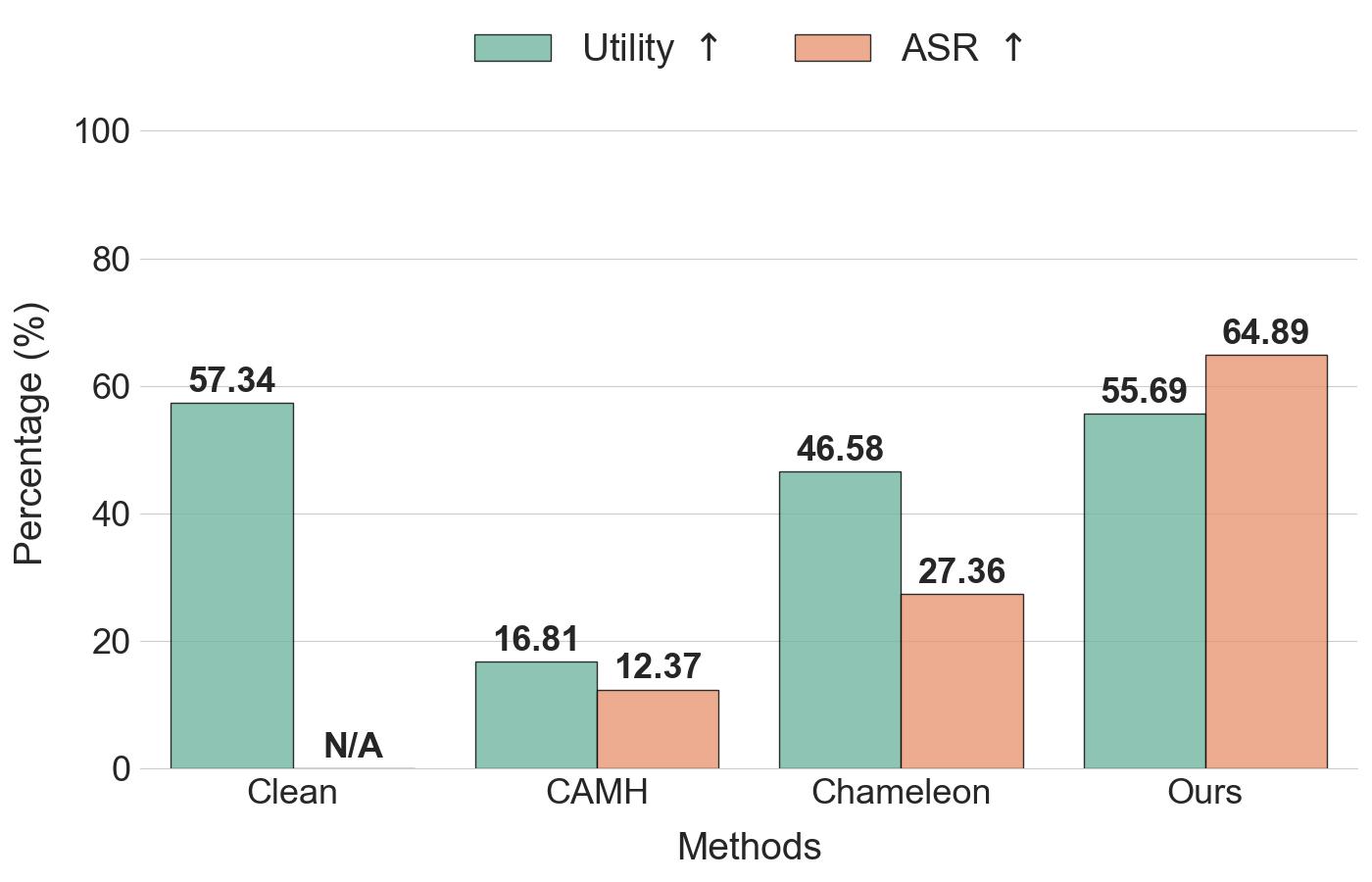}
        \caption{CIFAR-100, Tiny-ImageNet} 
        \label{fig:cham_cifar100_tiny_vgg}
    \end{subfigure}
    
    \caption{The results between the clean model, the CAMH \cite{DBLP:conf/aaai/HeCPLZWLJ25}, the Chameleon \cite{salem2022get} and Ours (approach under IPC $= 50$). The first row presents results using the ResNet18 architecture, while the second row displays results obtained with VGG16. Each figure is labeled in the sequence of the original dataset followed by the hijacking dataset. These results show that the OD attack preserves high hijacking performance even with limited samples, while delivering considerable utility.}
    \label{fig:baseline}
\end{figure*}

\begin{figure*}[!t]
    \centering
    \begin{subfigure}[b]{0.45\textwidth}
        \centering
        \includegraphics[width=\linewidth]{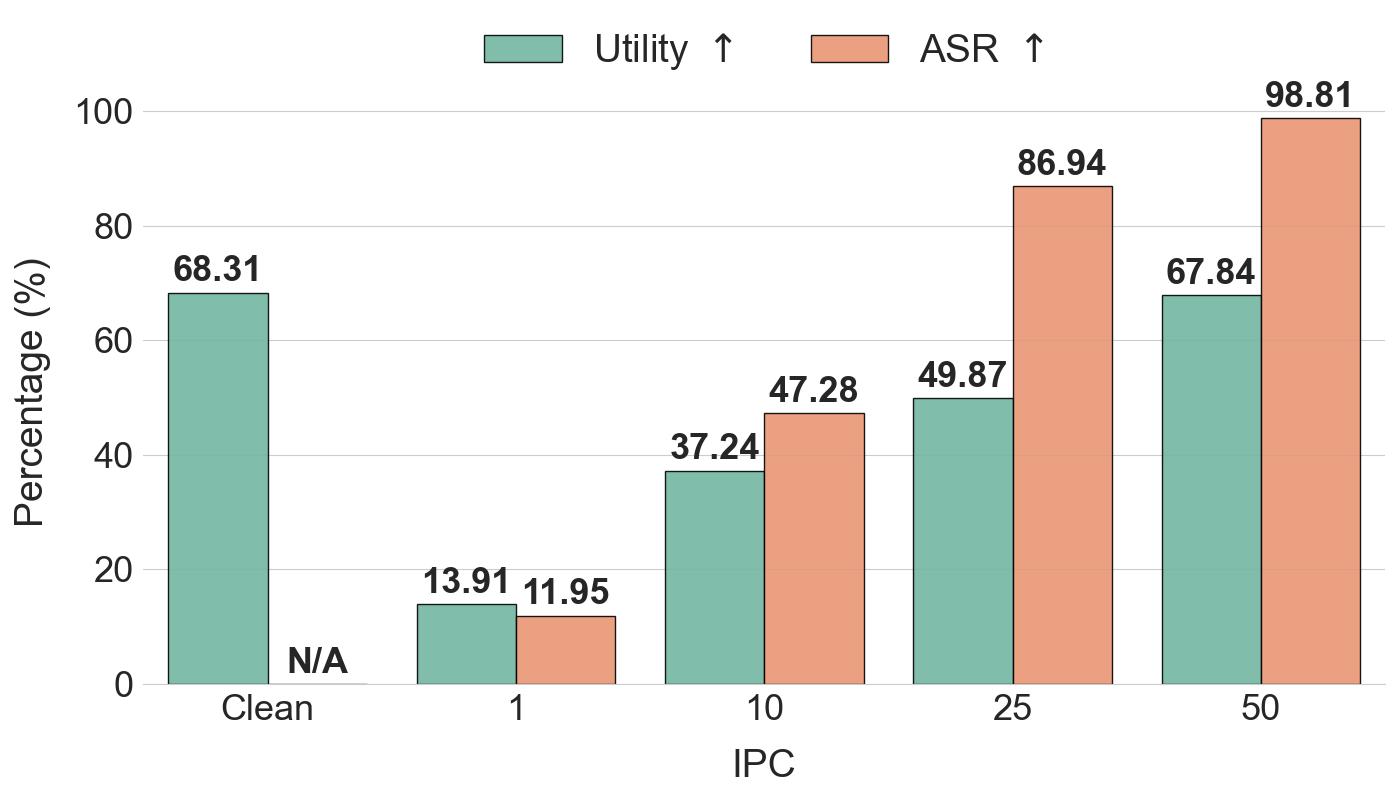}
        \caption{ResNet18: CIFAR-10, MNIST}
        \label{fig:ipc_cifar10_mnist_res18}
    \end{subfigure}
    \hspace{8pt}
    \begin{subfigure}[b]{0.45\textwidth}
        \centering
        \includegraphics[width=\linewidth]{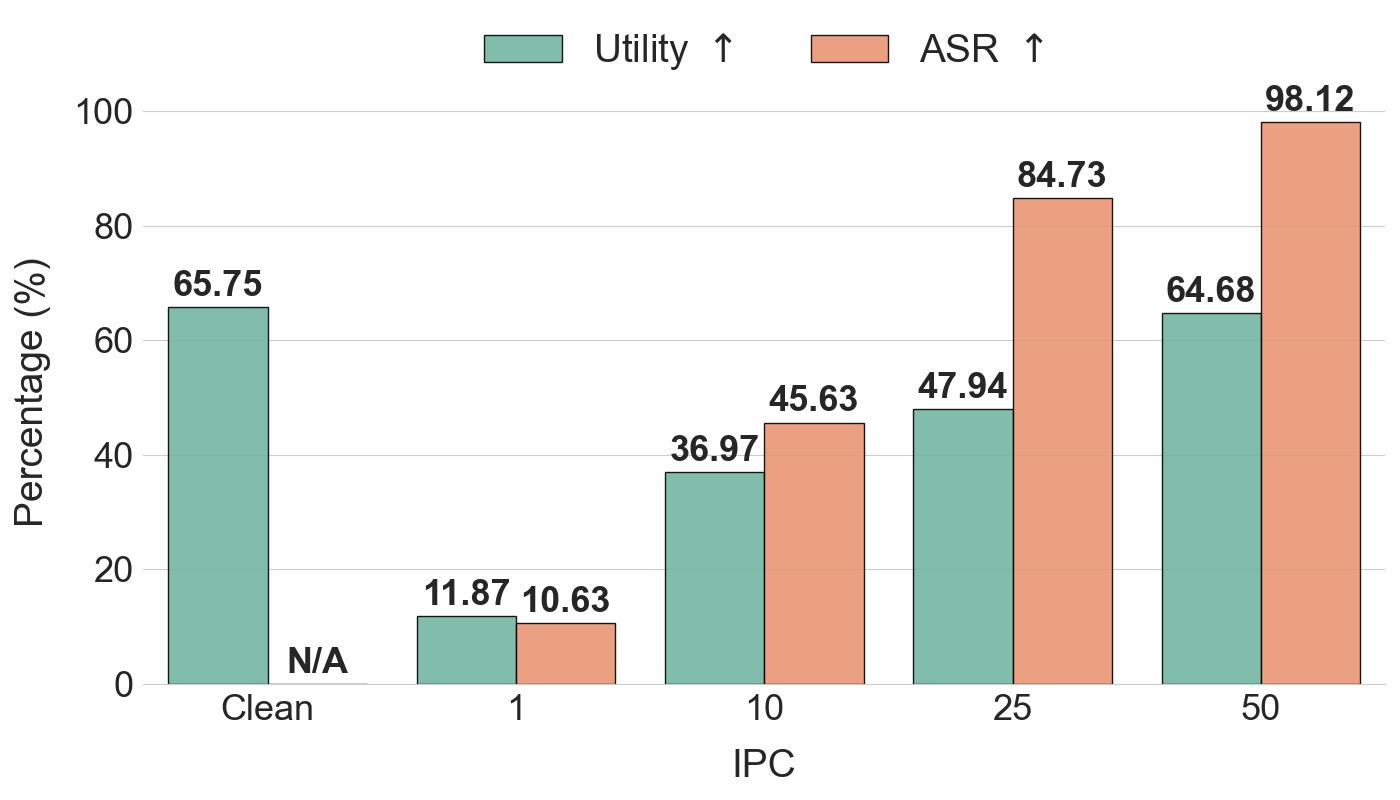}
        \caption{VGG16: CIFAR-10, MNIST}
        \label{fig:ipc_cifar10_mnist_vgg}
    \end{subfigure}
    
    \par\medskip 
    \begin{subfigure}[b]{0.45\textwidth}
        \centering
        \includegraphics[width=\linewidth]{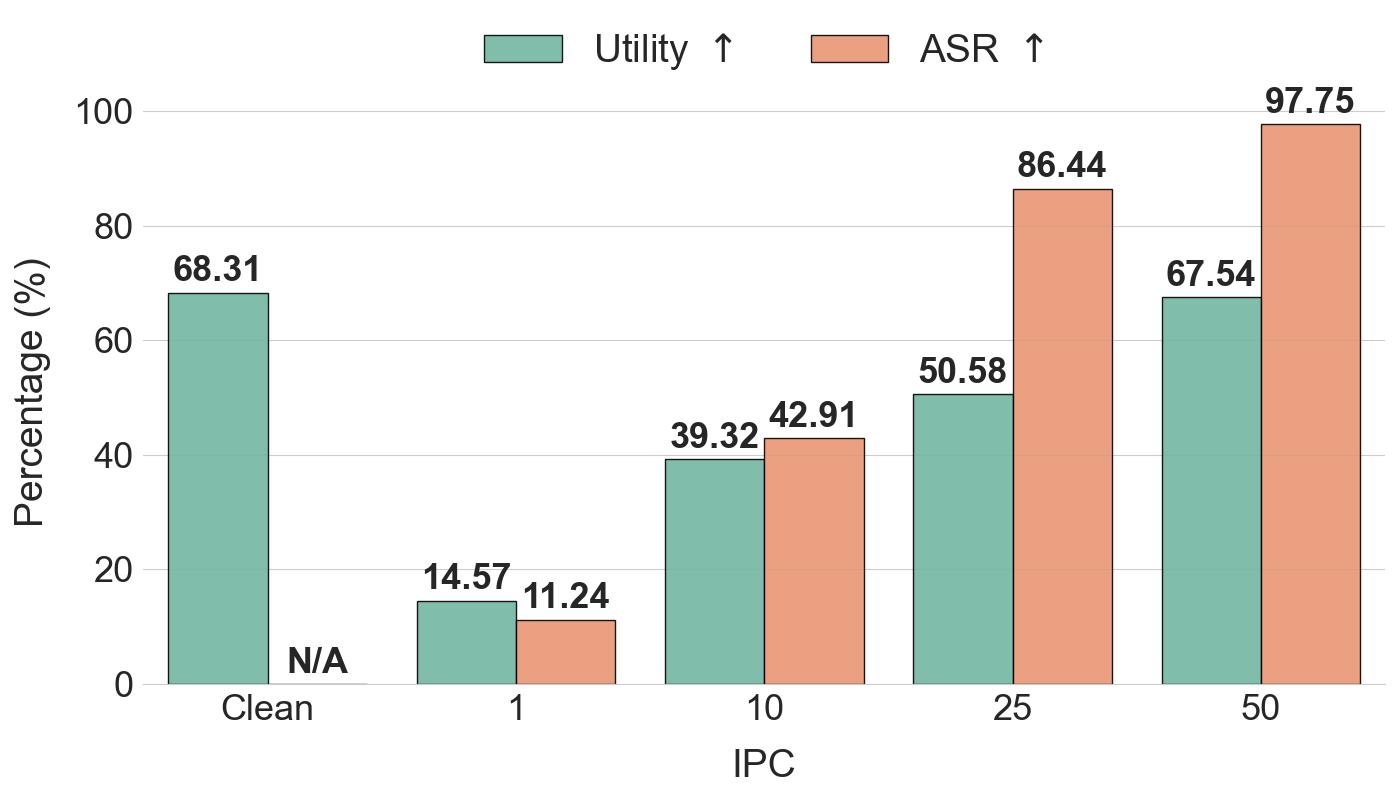}
        \caption{ResNet18: CIFAR-10, SVHN}
        \label{fig:ipc_cifar10_svhn_res18}
    \end{subfigure}
    \hspace{8pt}
    \begin{subfigure}[b]{0.45\textwidth}
        \centering
        \includegraphics[width=\linewidth]{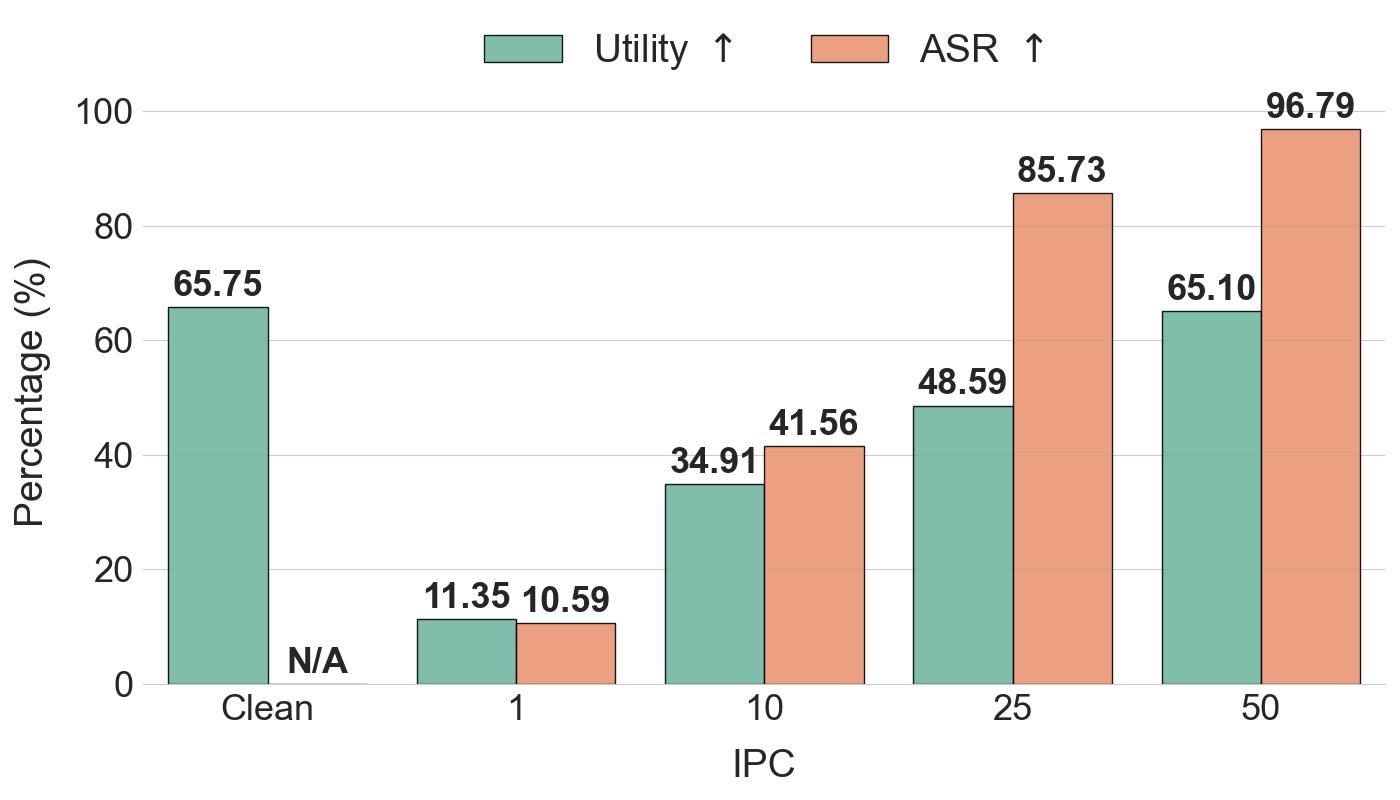}
        \caption{VGG16: CIFAR-10, SVHN}
        \label{fig:ipc_cifar10_svhn_vgg}
    \end{subfigure}
    
    \par\medskip 
    \begin{subfigure}[b]{0.45\textwidth}
        \centering
        \includegraphics[width=\linewidth]{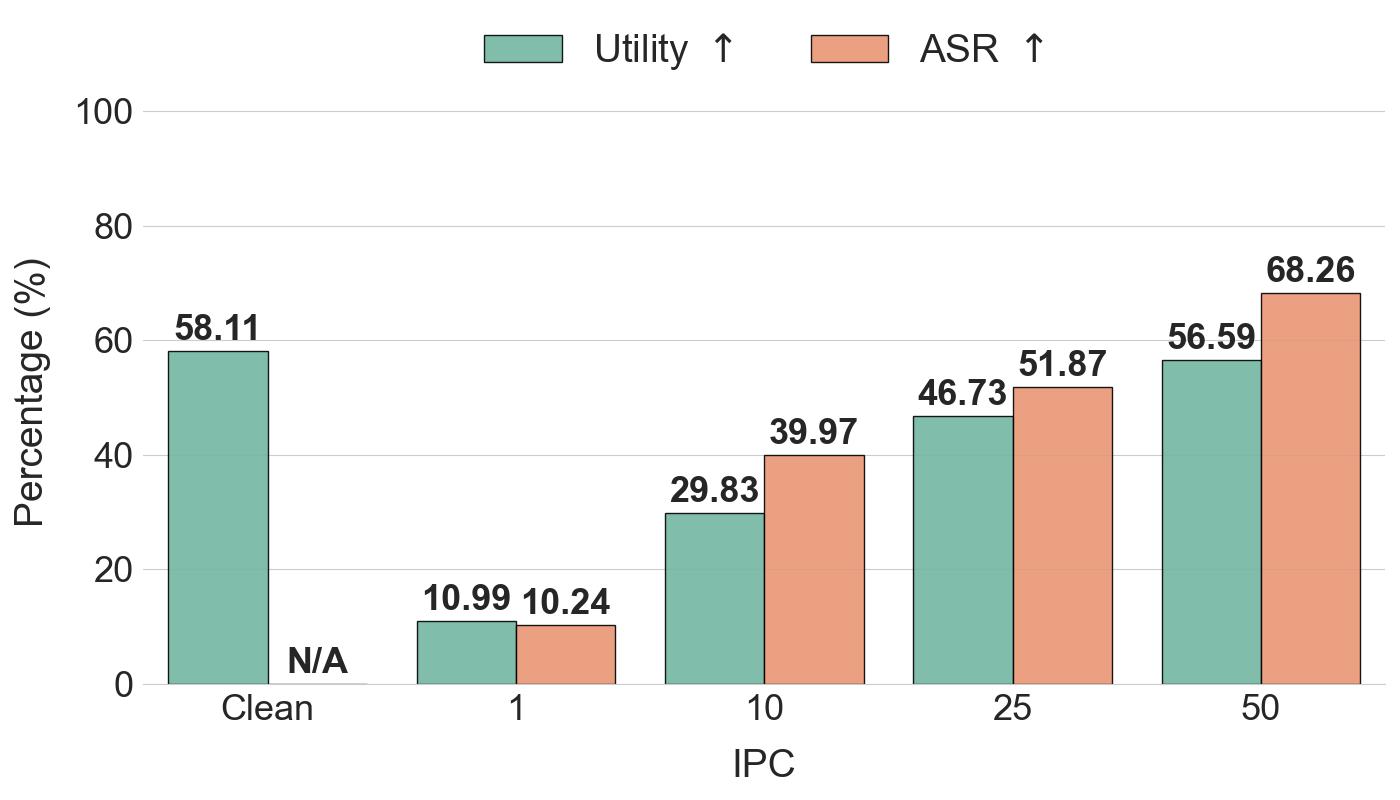}
        \caption{ResNet18: CIFAR-100, Tiny-ImageNet}
        \label{fig:ipc_cifar100_tiny_res18}
    \end{subfigure}
    \hspace{8pt}
    \begin{subfigure}[b]{0.45\textwidth}
        \centering
        \includegraphics[width=\linewidth]{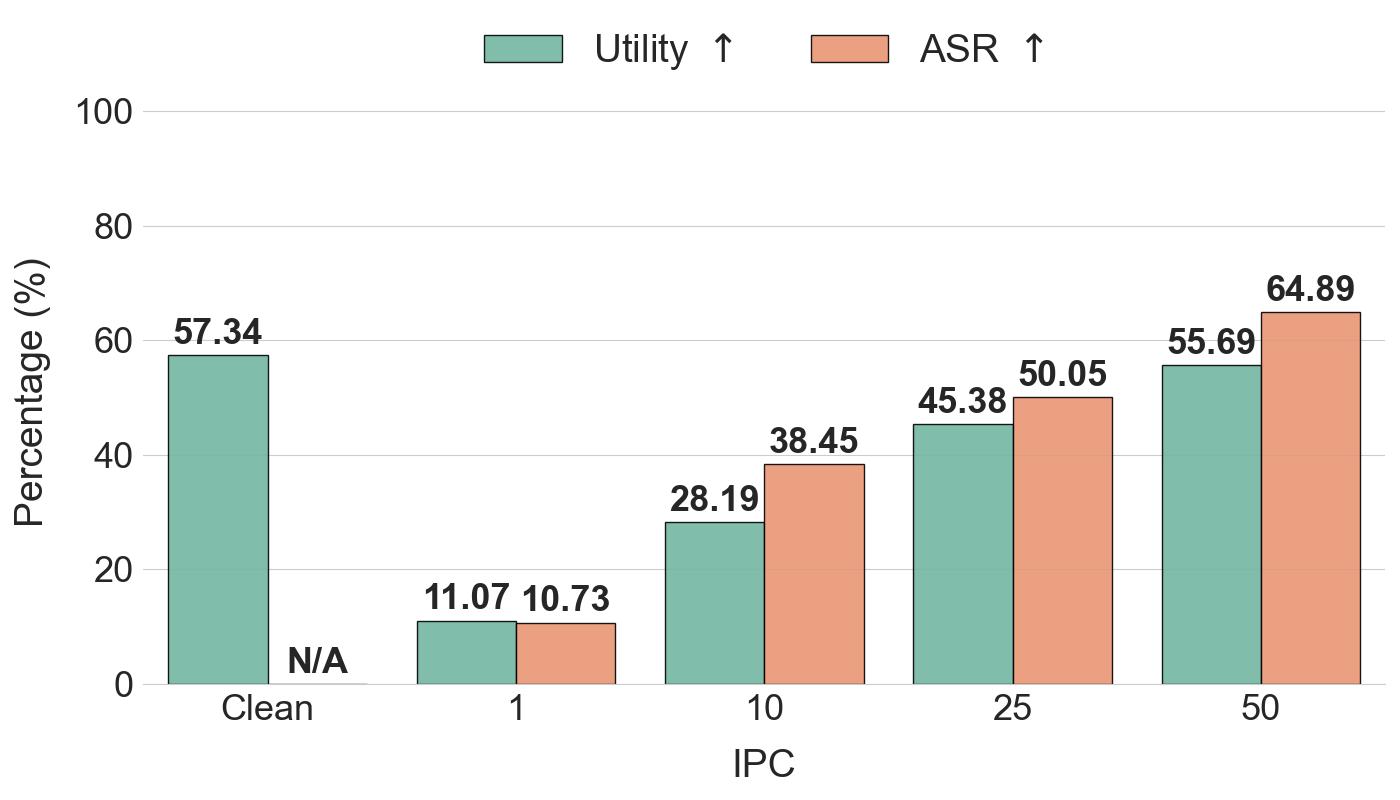}
        \caption{VGG16: CIFAR-100, Tiny-ImageNet}
        \label{fig:ipc_cifar100_tiny_vgg}
    \end{subfigure}

    \par\medskip
    \begin{subfigure}[b]{0.45\textwidth}
        \centering
        \includegraphics[width=\linewidth]{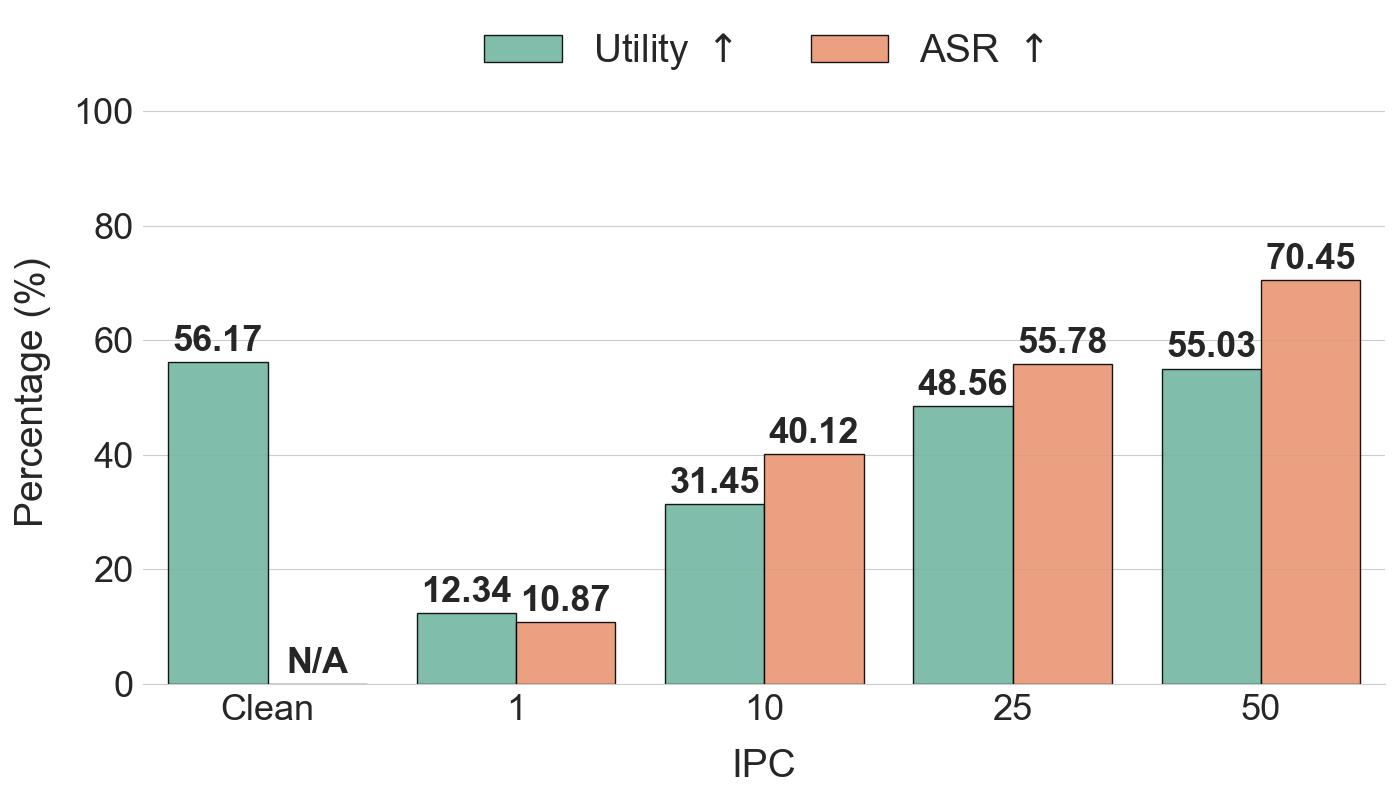}
        \caption{ResNet18: ImageNet}
        \label{fig:ipc_image_image_res18}
    \end{subfigure}
    \hspace{8pt}
    \begin{subfigure}[b]{0.45\textwidth}
        \centering
        \includegraphics[width=\linewidth]{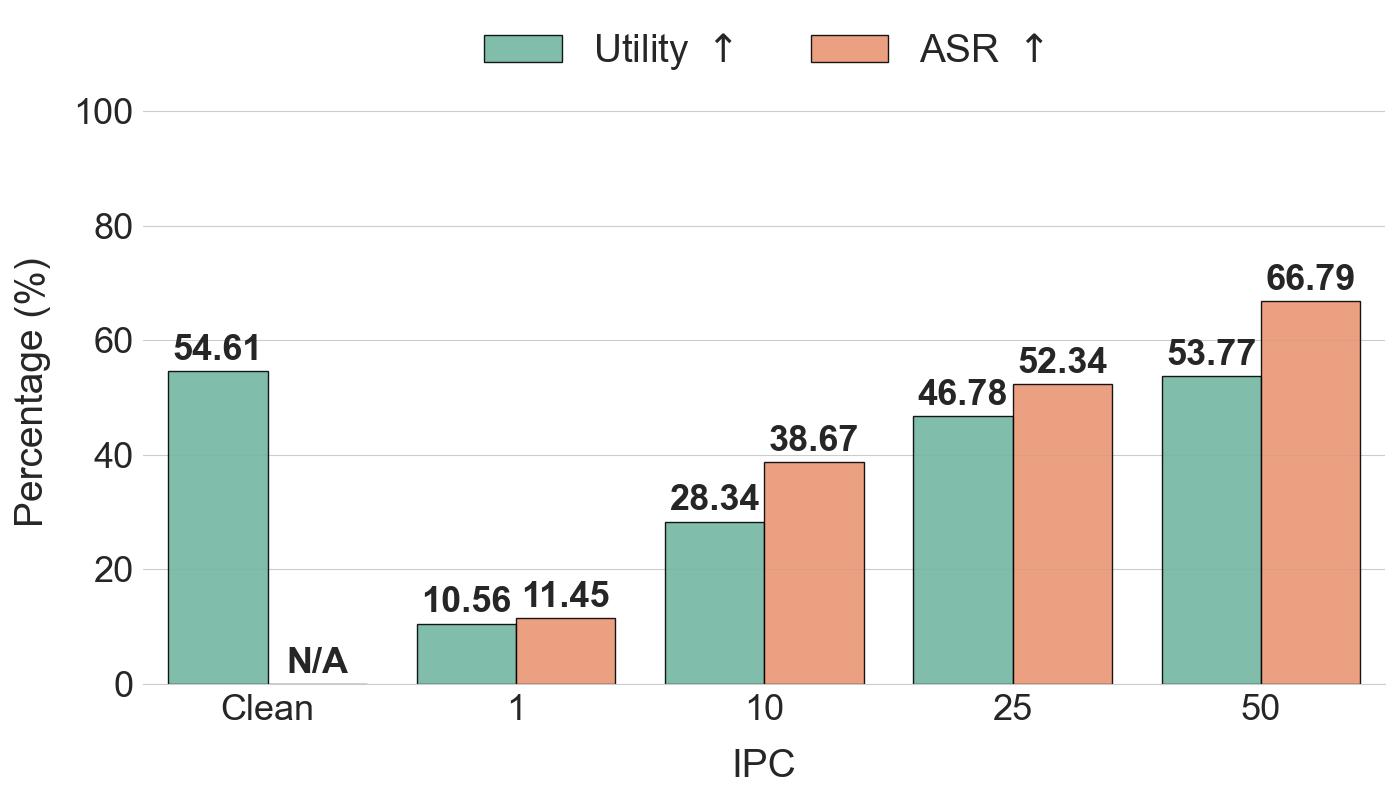}
        \caption{VGG16: ImageNet}
        \label{fig:ipc_image_image_vgg}
    \end{subfigure}
    
    \caption{Evaluation of Different IPC of OD attack. The left column displays results using the \textbf{ResNet18} architecture, while the right column displays results obtained with \textbf{VGG16}. The rows correspond to different dataset pairs.}
    \label{fig:IPC}
\end{figure*}

\subsection{Performance Evaluation}
\subsubsection{Effectiveness of OD}
In this section, we compare OD attack with a clean model, serving as a control group for utility metrics; and two SOTA baselines based on the CAMH and Chameleon attack methods.
To ensure a fair comparison, we set the training data volume for CAMH and Chameleon attacks to $50\%$, simulating attack scenarios with limited training data.
Moreover, for the clean model trained on the distilled dataset and our proposed OD attack, we restrict the number of images per class (IPC) to 50. 
This setup aims to demonstrate the performance of the OD attack under a regime of extremely limited training samples.

As shown in the results of \autoref{fig:baseline}, our model utility is comparable to that of the clean model across all datasets and model architectures, with a maximum discrepancy of only $1.52\%$.
Although, both the CAMH and Chameleon attacks suffer from significant degradation in model utility.
Nevertheless, Chameleon exhibits higher utility than CAMH because it incorporates the benign original dataset during training.
This comparative analysis shows that the OD attack achieves superior performance in terms of model utility.
Consequently, victims cannot detect the presence of malicious tasks within the dataset based on variations in model utility, thereby confirming the high stealthiness of the OD attack.

Furthermore, the ASR of the OD attack consistently surpasses that of the baseline methods.
Specifically, for all 10-class tasks, the OD attack achieves an ASR exceeding $96\%$, while for 100-class tasks, the ASR remains above $64\%$.
These results demonstrate the strong efficacy of the OD attack in scenarios with extremely limited training samples.
Although the ASR for 100-class tasks shows a decline compared to that of 10-class tasks, we attribute this drop to the inherent limitations of current distillation methods in handling tasks with 100 or more classes.
    
\subsubsection{Impact of IPC}
To verify whether the OD attack can effectively reduce the number of required osmosis samples, we set the IPC to 1, 10, 25 and 50, respectively.
For the clean model, the distilled dataset is configured with an IPC of 50.
The left column of \autoref{fig:IPC} illustrates the performance of the OD attack on ResNet18, while the right column displays the performance on VGG16.
The comparison reveals that across all 10-class tasks (\Cref{fig:ipc_cifar10_mnist_res18,fig:ipc_cifar10_svhn_res18,fig:ipc_cifar10_mnist_vgg,fig:ipc_cifar10_svhn_vgg}) with an IPC of 50, the OD attack achieves model utility comparable to that of the clean model, with an ASR exceeding $96\%$.
When the IPC reduced to 25, model utility declines, yet the ASR remains above $84\%$.
For 100-class tasks (\Cref{fig:ipc_cifar100_tiny_res18,fig:ipc_cifar100_tiny_vgg,fig:ipc_image_image_res18,fig:ipc_image_image_vgg}), the OD attack continues to yield model utility comparable to the clean model. 
Even when both the hijacking and original dataset consist of high-resolution images, the OD attack maintains model utility comparable to the clean baseline and an ASR above $65\%$.
Notably, the ASR for hijacking tasks using the ImageNet dataset is slightly higher than those using Tiny-ImageNet; we attribute this to the fact that both the ResNet18 and VGG16 pre-trained models were originally trained on ImageNet.
In summary, the OD attack demonstrates high utility for the original task and a high success rate for the hijacking task.
Moreover, it maintains effective ASR and utility across different victim models, highlighting its robustness and generalizability.

\subsubsection{Impact of Dataset Correlation}
To investigate the impact of dataset differences on the OD attack, we designed two groups of experiments. 
The first group is set under the condition where the original dataset and the hijacking dataset are unrelated, using CIFAR-10 as original task and SVHN as the hijacking task.
The second group is set under the condition where the original dataset and the hijacking dataset are related, using CIFAR-100 as the original task and CIFAR-10 as the hijacking task.
Moreover, both groups are set to IPC $= 50$. 

\begin{figure}[!t]
    \centering
    \includegraphics[width=1.0\linewidth]{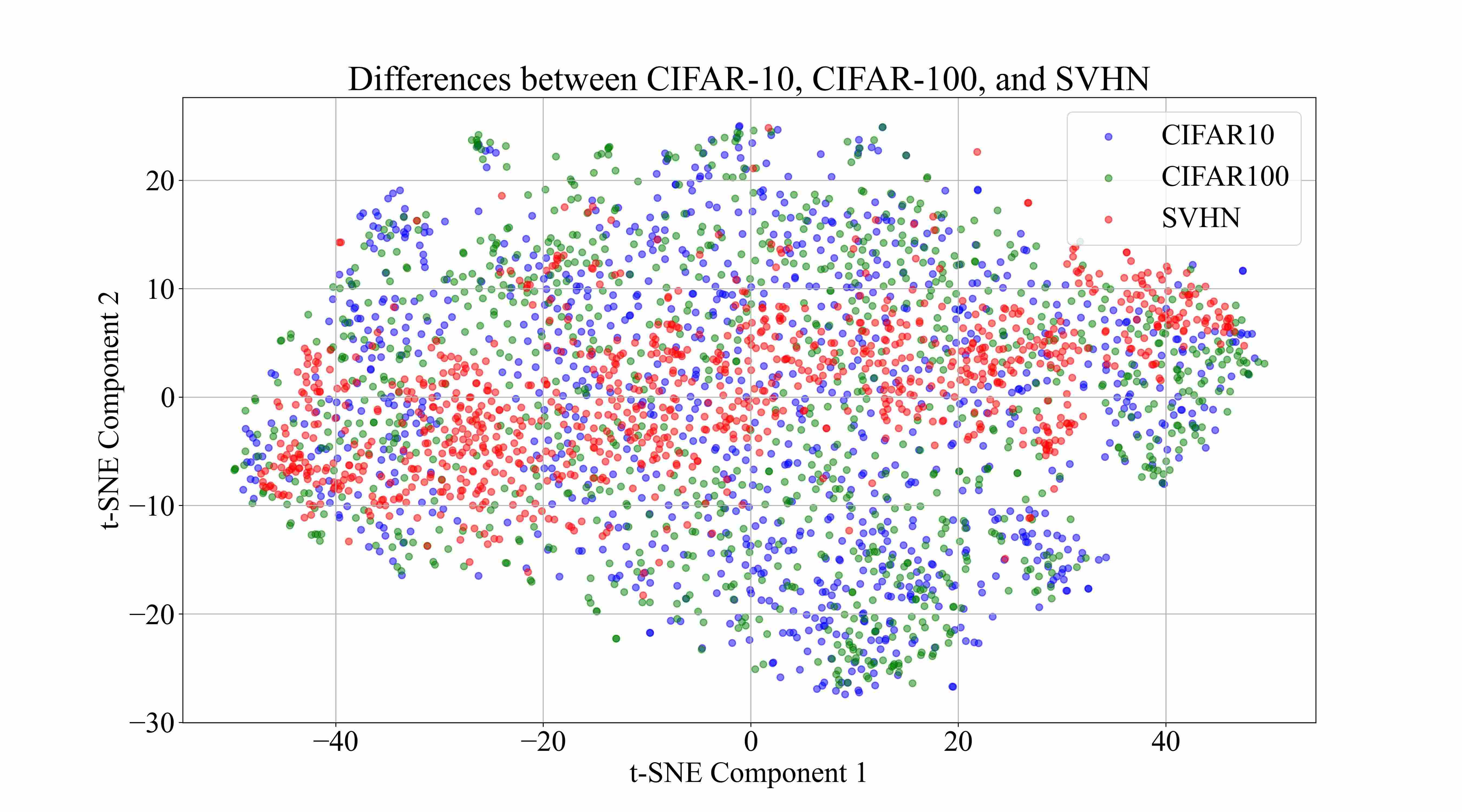}
    \caption{Visualization of comparing different datasets using t-distributed Stochastic Neighbor Embedding (t-SNE) \cite{JMLR:v9:vandermaaten08a}.}
    \label{fig:t-SNE}
\end{figure}

\begin{figure}[!t]
    \centering
    \begin{subfigure}[b]{0.48\textwidth}
        \centering
        \includegraphics[width=\linewidth]{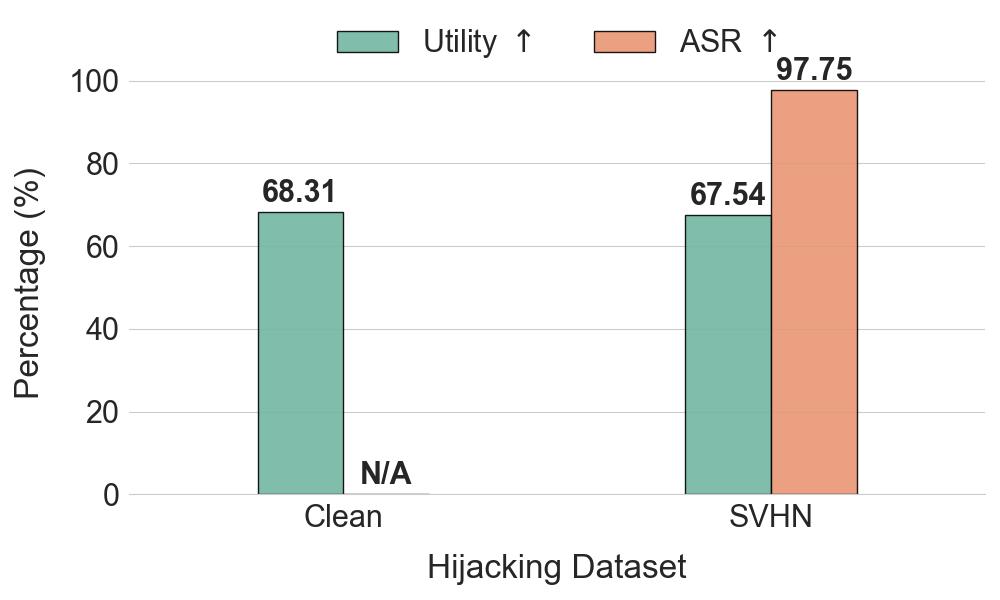}
        \caption{CIFAR-10, SVHN}
        \label{fig:correlation_cifar-10-svhn}
    \end{subfigure}
    \hfill
    \begin{subfigure}[b]{0.48\textwidth}
        \centering
        \includegraphics[width=\linewidth]{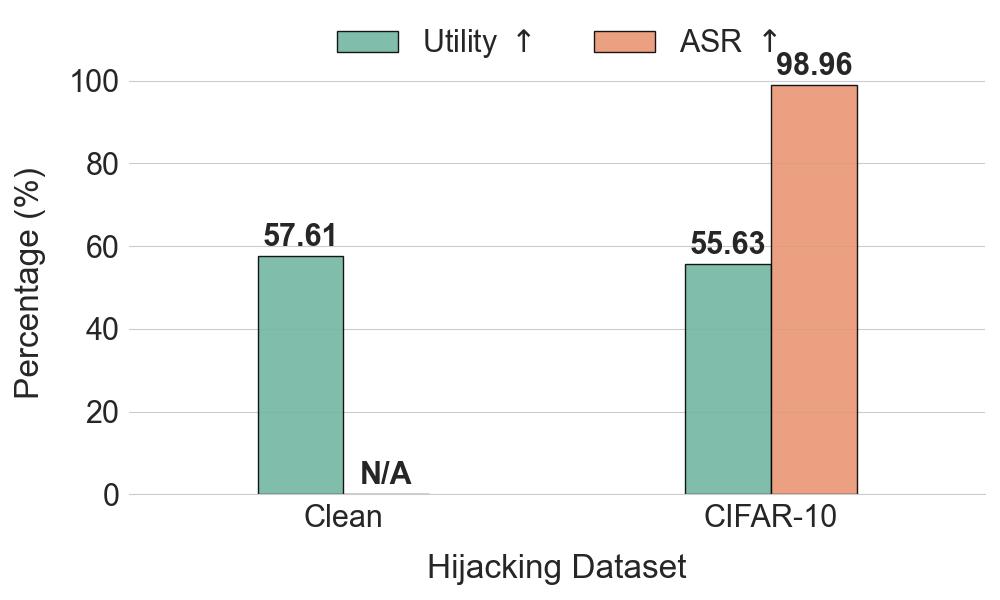}
        \caption{CIFAR-100, CIFAR-10}
        \label{fig:correlation_cifar-100-cifar-10}
    \end{subfigure}
     \caption{Experimental results on dataset correlation.}
    \label{fig: comparison}
\end{figure}

To provide a more intuitive illustration of the differences among the CIFAR-10, CIFAR-100 and SVHN datasets, we visualize the distribution variations across these datasets using t-SNE (as shown in \autoref{fig:t-SNE}). 
It is evident that, in the first group of experiments, SVHN and CIFAR-10 exhibit larger distribution differences, whereas in the second group, CIFAR-10 and CIFAR-100 show similar distributions.
Furthermore, \autoref{fig: comparison} presents the results under the two different settings. 
\autoref{fig:correlation_cifar-10-svhn} corresponds to the case where the datasets are unrelated, while \autoref{fig:correlation_cifar-100-cifar-10} corresponds to the case where the datasets are related.
In both groups the ASR exceeds $97\%$, and the utility is comparable to that of the clean model.
Notably, the ASR in \autoref{fig:correlation_cifar-100-cifar-10} is slightly higher than \autoref{fig:correlation_cifar-10-svhn}, which is attributed to the high similarity between the CIFAR-10 and CIFAR-100 datasets. 
Through our experiments, we demonstrate that the OD attack exhibits strong attack performance regardless of whether the hijacking dataset is related to the original dataset, highlighting the generalization capability of the OD attack.

\subsection{Stealthiness Analysis}
\begin{figure}[!t]
    \centering
    \begin{subfigure}[b]{0.48\textwidth}
        \centering
        \includegraphics[width=\linewidth]{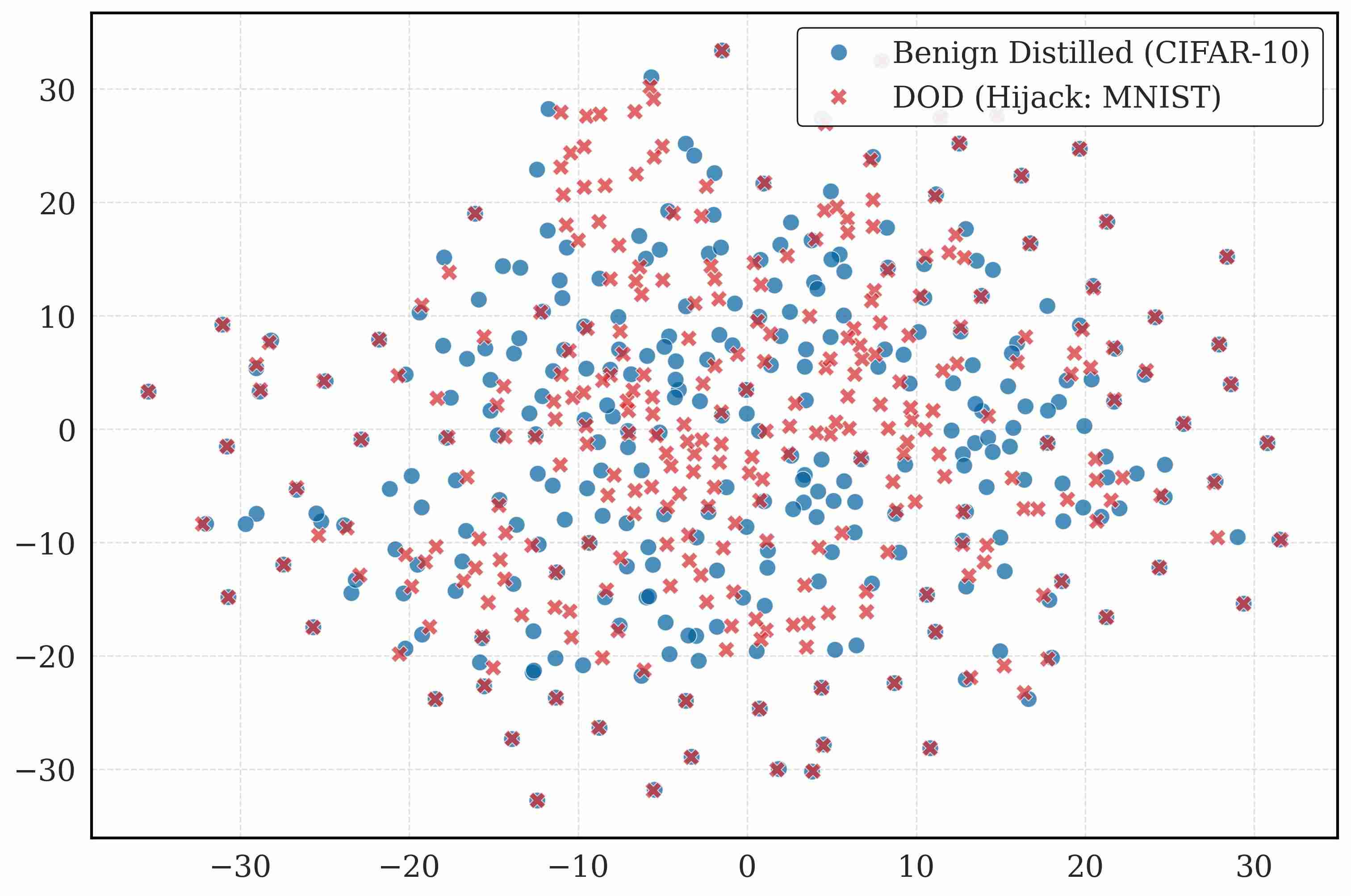}
        \caption{CIFAR-10, MNIST}
        \label{fig:t-sne_DOD-MNIST}
    \end{subfigure}
    \hfill
    \begin{subfigure}[b]{0.48\textwidth}
        \centering
        \includegraphics[width=\linewidth]{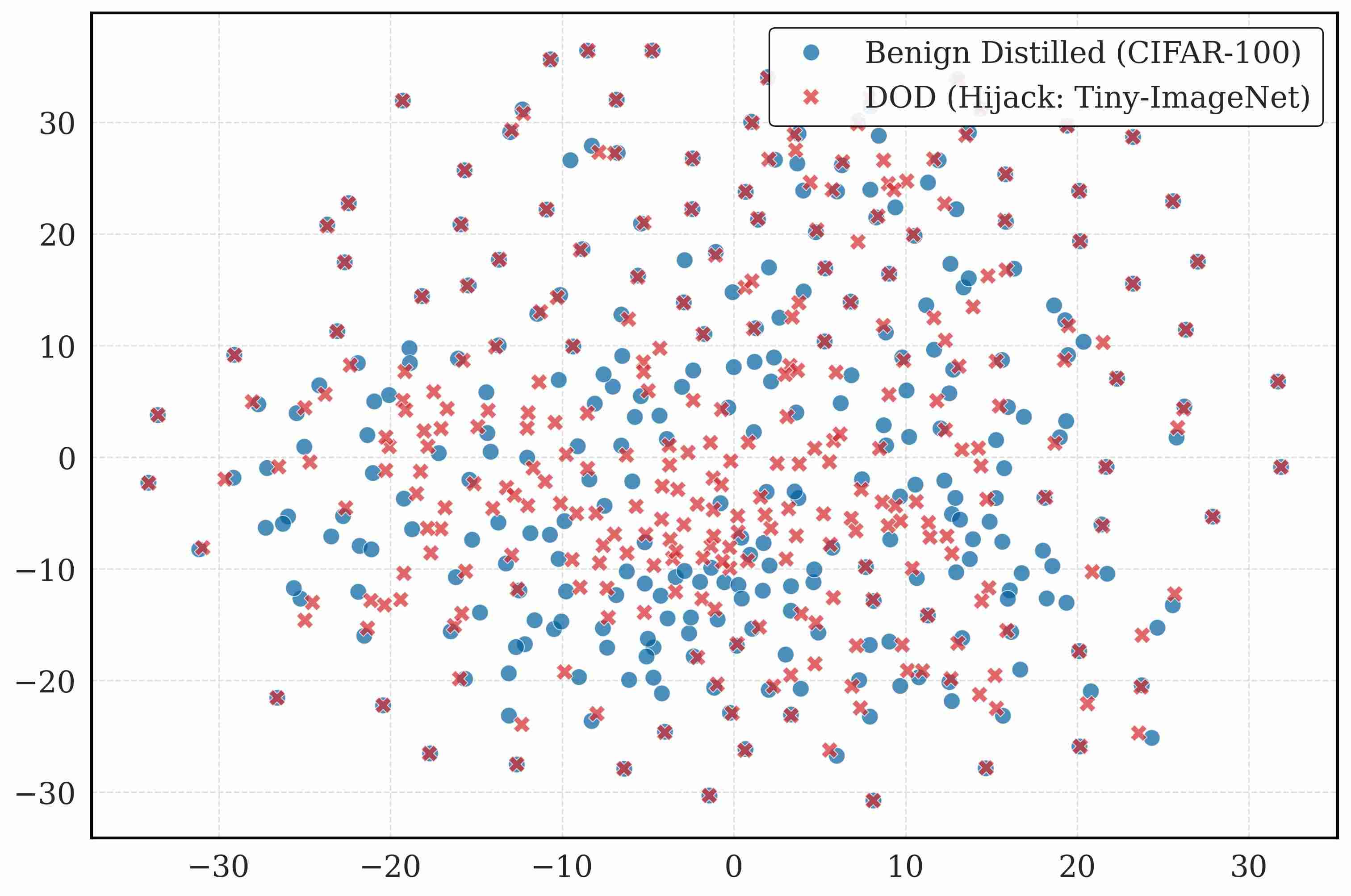}
        \caption{CIFAR-100, Tiny-ImageNet}
        \label{fig:t-SNE_DOD-cifar-100}
    \end{subfigure}
     \caption{Visualization comparison of benign distillation dataset and DOD using t-SNE.}
     \label{fig:t-SNE of DOD and Benign Distilled}
\end{figure}

To further evaluate the stealthiness of the OD attack, we extracted feature vectors from the penultimate layer of the victim model and visualized both the benign distilled dataset and the DOD using t-SNE dimensionality reduction.
For the 10-class task, we selected MNIST as the hijacking dataset and CIFAR-10 as the original dataset; similarly, for the 100-class task, we used Tiny-ImageNet as the hijacking dataset and CIFAR-100 as the original dataset.
As shown in \autoref{fig:t-SNE of DOD and Benign Distilled}, for both 10-class and 100-class tasks, samples from the benign distilled dataset and the DOD are highly intermingled in the feature space, forming no distinct clusters.
This indicates that the Transporter architecture successfully embeds the features of the malicious tasks into the manifold of the benign data.
This proves that the distilled osmosis samples are indistinguishable from the benign samples at the feature level, thereby confirming the robust stealthiness of the OD attack.
Consequently, it is difficult for victims to detect the presence of malicious tasks based on feature analysis.

\begin{figure*}[!t]
    \centering
    \begin{subfigure}[b]{0.42\textwidth}
        \centering
        \includegraphics[width=\linewidth]{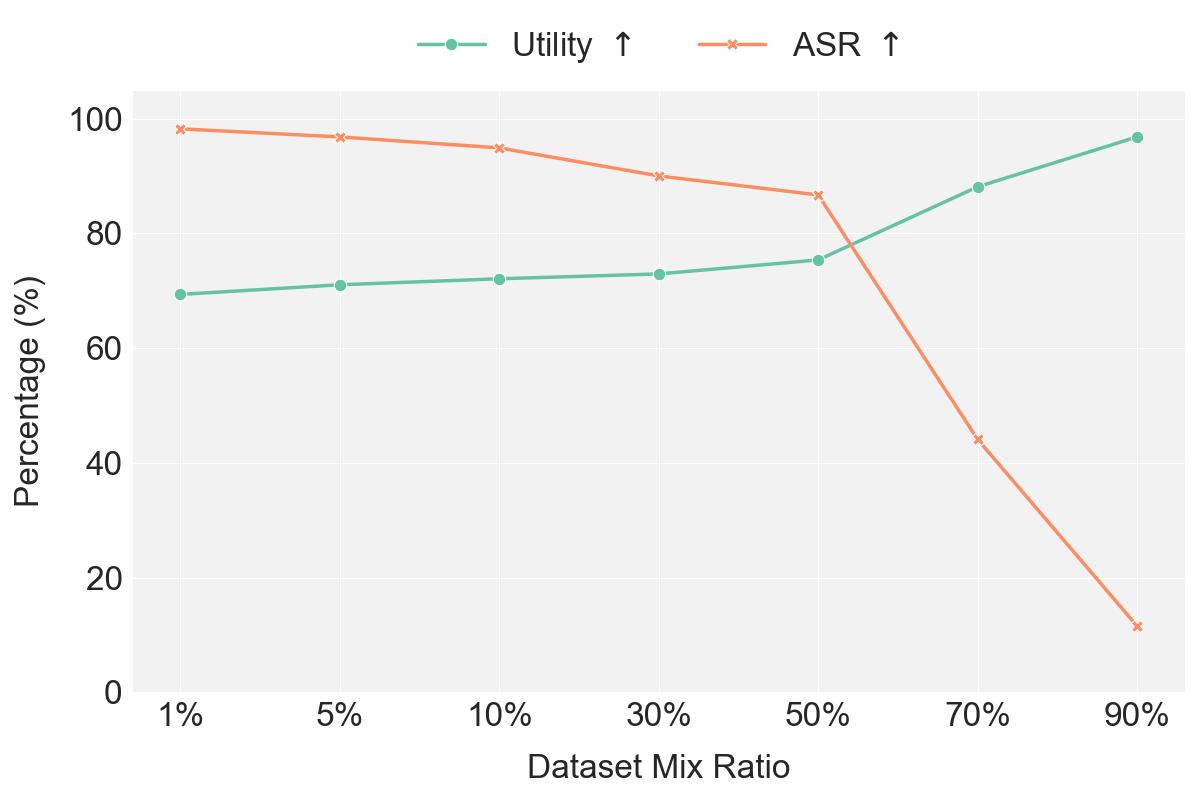}
        \caption{ResNet18: CIFAR-10, MNIST}
        \label{fig:mix_cifar10_mnist_res18}
    \end{subfigure}
    \hspace{8pt}
    \begin{subfigure}[b]{0.42\textwidth}
        \centering
        \includegraphics[width=\linewidth]{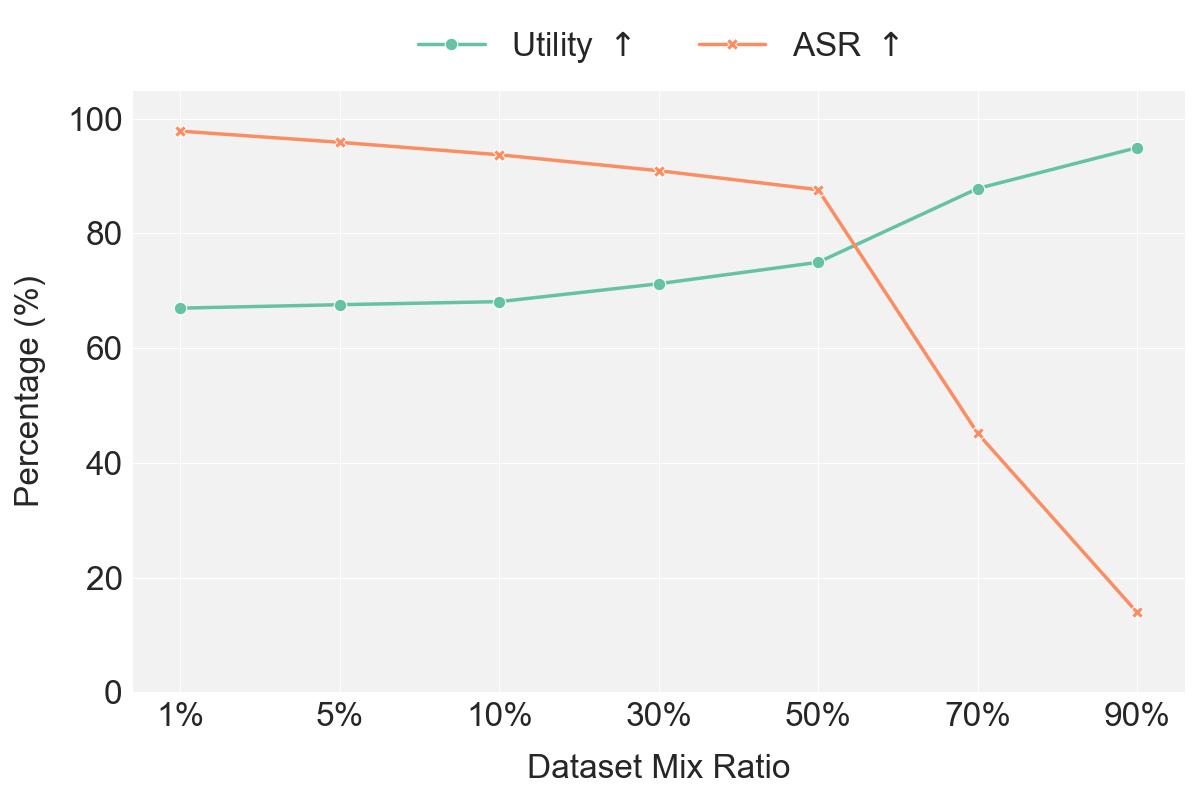}
        \caption{VGG16: CIFAR-10, MNIST}
        \label{fig:mix_cifar10_mnist_vgg}
    \end{subfigure}
    
    \par\medskip 
    \begin{subfigure}[b]{0.42\textwidth}
        \centering
        \includegraphics[width=\linewidth]{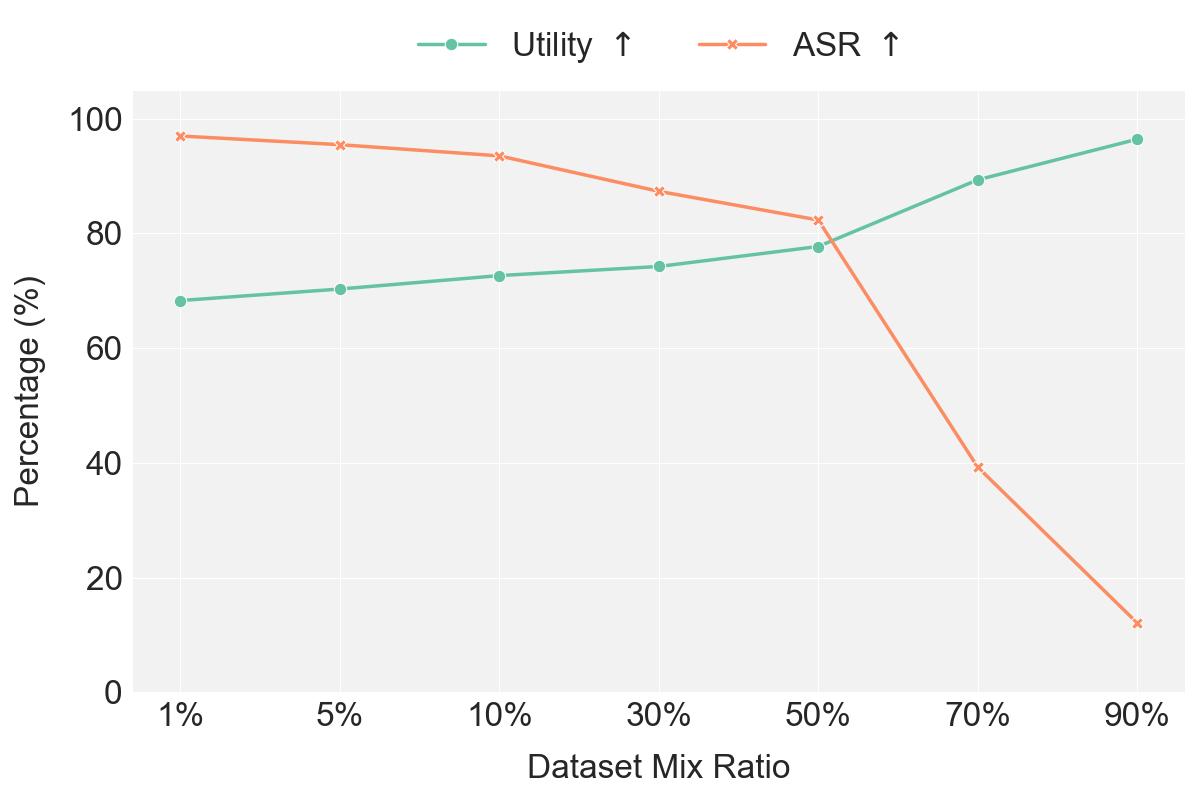}
        \caption{ResNet18: CIFAR-10, SVHN}
        \label{fig:mix_cifar10_svhn_res18}
    \end{subfigure}
    \hspace{8pt}
    \begin{subfigure}[b]{0.42\textwidth}
        \centering
        \includegraphics[width=\linewidth]{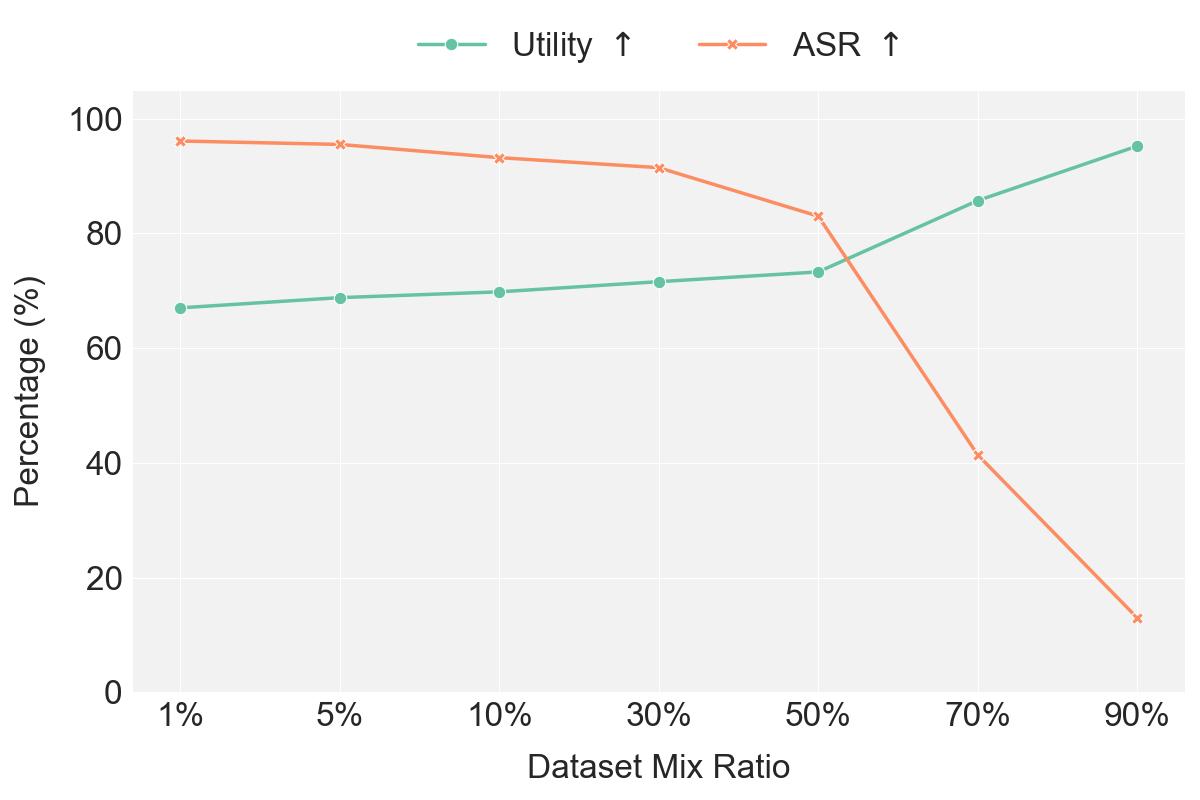}
        \caption{VGG16: CIFAR-10, SVHN}
        \label{fig:mix_cifar10_svhn_vgg}
    \end{subfigure}
    
    \par\medskip 
    \begin{subfigure}[b]{0.42\textwidth}
        \centering
        \includegraphics[width=\linewidth]{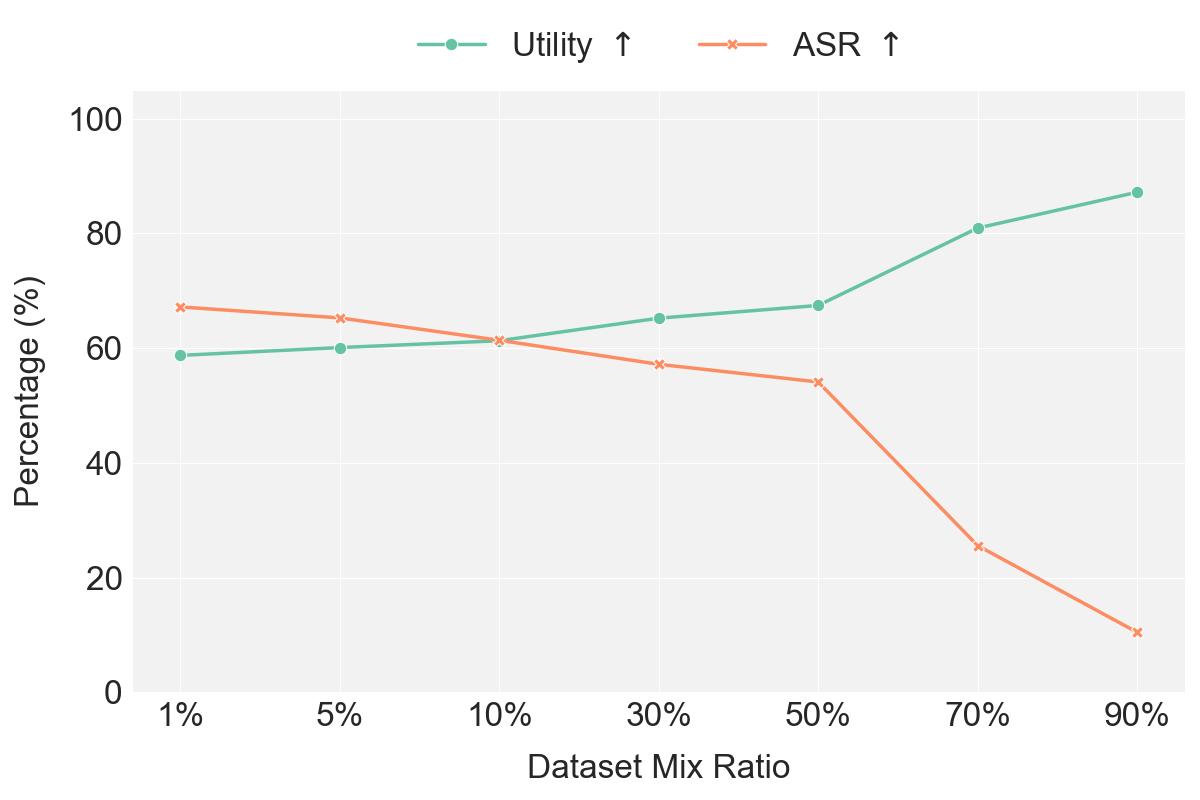}
        \caption{ResNet18: CIFAR-100, Tiny-ImageNet}
        \label{fig:mix_cifar100_tiny_res18}
    \end{subfigure}
    \hspace{8pt}
    \begin{subfigure}[b]{0.42\textwidth}
        \centering
        \includegraphics[width=\linewidth]{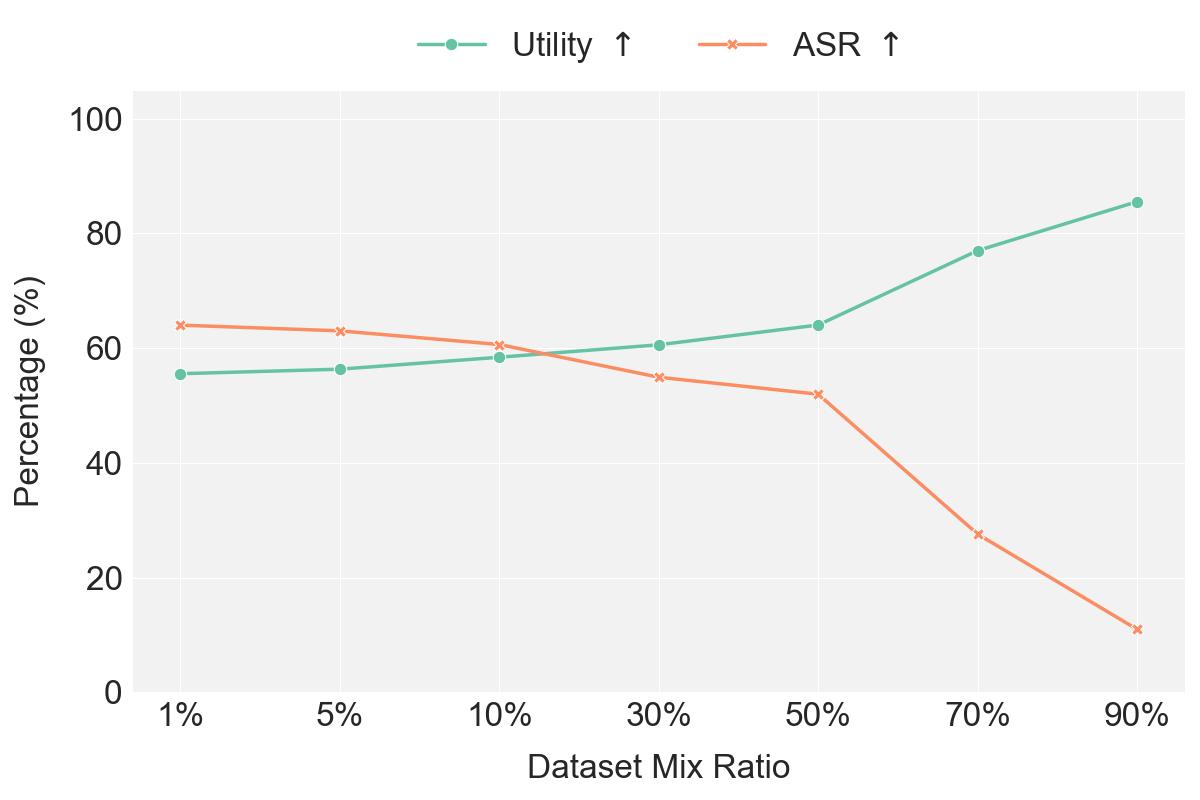}
        \caption{VGG16: CIFAR-100, Tiny-ImageNet}
        \label{fig:mix_cifar100_tiny_vgg}
    \end{subfigure}

    \par\medskip
    \begin{subfigure}[b]{0.42\textwidth}
        \centering
        \includegraphics[width=\linewidth]{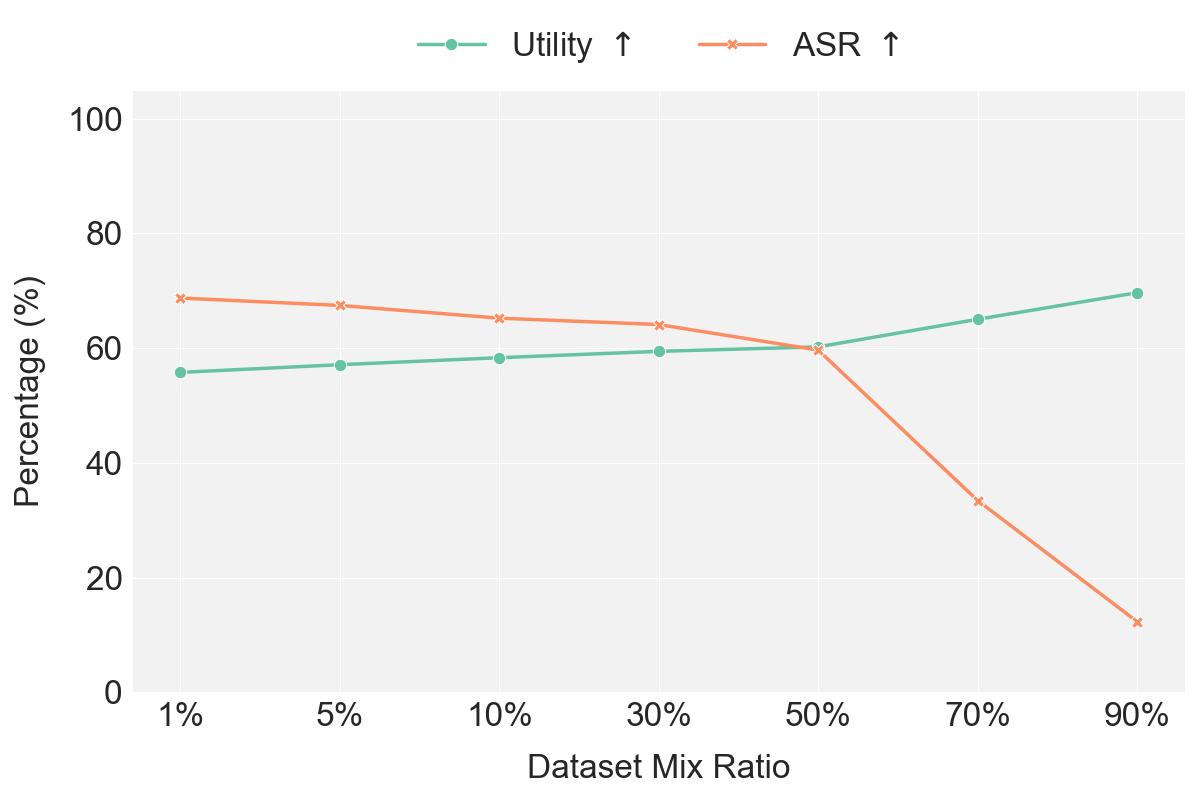}
        \caption{ResNet18: ImageNet}
        \label{fig:mix_image_image_res18}
    \end{subfigure}
    \hspace{8pt}
    \begin{subfigure}[b]{0.42\textwidth}
        \centering
        \includegraphics[width=\linewidth]{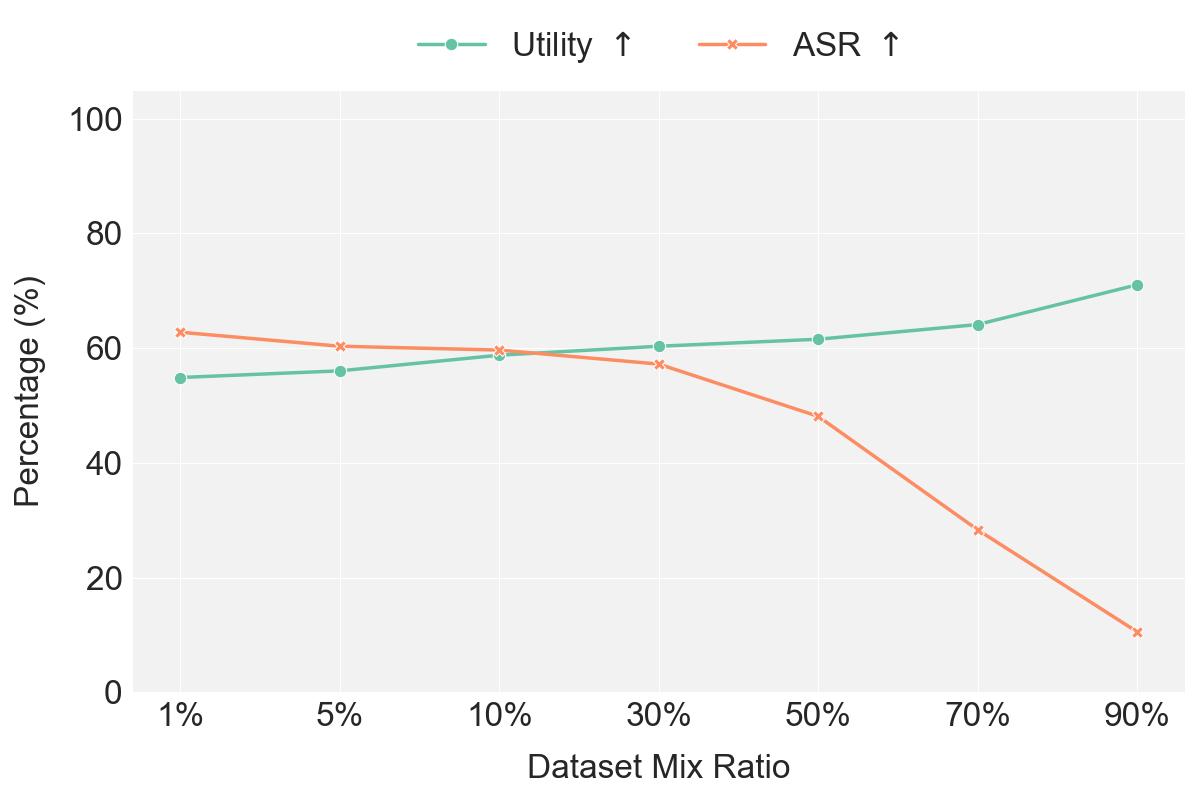}
        \caption{VGG16: ImageNet}
        \label{fig:mix_image_image_vgg}
    \end{subfigure}
    \caption{Experimental results on data dilution robustness.}
    \label{fig:Mix ratio}
\end{figure*}

\begin{figure}[!t]
    \centering
    \includegraphics[width=1.0\linewidth]{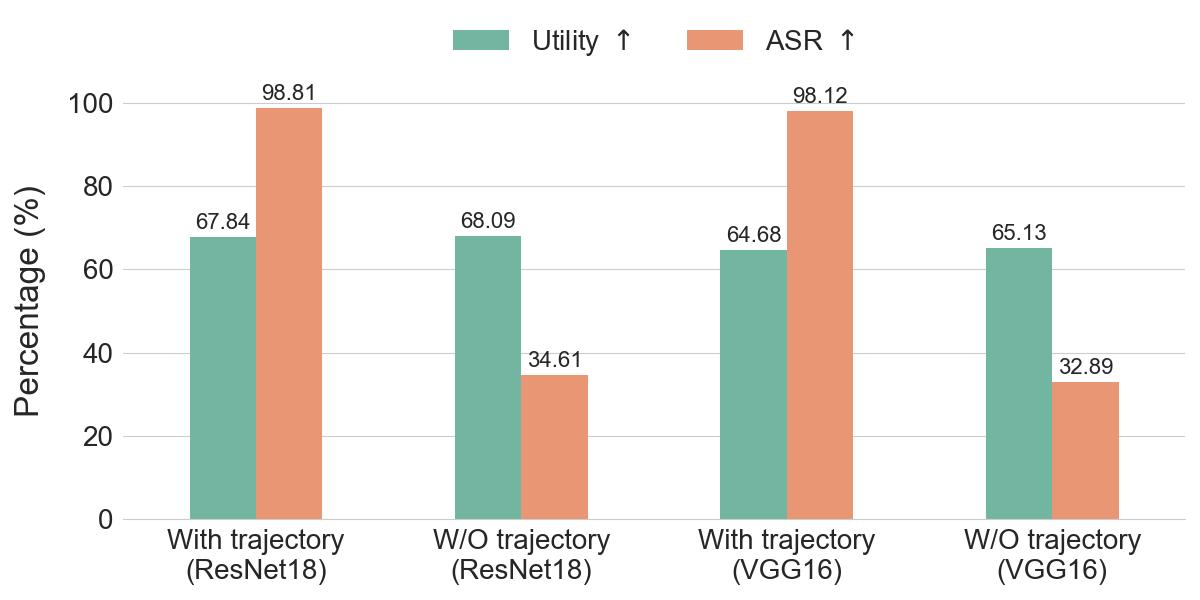}
    \caption{The results of the ablation study on whether training trajectory matching is used during the distillation stage, with CIFAR-10 as the original dataset, MNIST as the hijacking dataset, and IPC $= 50$.}
    \label{fig: mtt}
\end{figure}

\subsection{Ablation Study}
\subsubsection{Impact of the Trajectory Loss}
During the distillation process, we employed a training trajectory matching method to ensure that the distilled osmosis samples retained the features of the hijacking samples.
To verify the necessity of incorporating this training trajectory matching approach and its potential to enhance the performance of the OD attack, we conducted ablation experiments.
In these experiments, we selected CIFAR-10 as the original dataset and MNIST as the hijacking dataset. The results in \autoref{fig: mtt} clearly demonstrate that the ASR of the model trained with samples that have the training trajectory information is significantly higher than that of the model trained without it.
This indicates that adopting training trajectory matching is crucial for the OD attack.
Furthermore, as shown in \autoref{fig: mtt}, incorporating training trajectory matching does not influence the model's utility, suggesting that our attack imposes negligible effects on the original task while demonstrating exceptional stealth capabilities.

\subsection{Robustness Evaluation}
\subsubsection{Impact of Key Patches}
To illustrate the impact of the number of key patches spliced within a single synthesized image ($N$) on utility and ASR, we conducted experiments using varying numbers of key patches ($N \in \{1,4,9,16,25\}$) under the setting of $IPC=50$.
As shown in \autoref{fig:Key Patch}, when $N=1$, the distillation stage selects only a single patch based on the realism score without splicing other patches; in this case, both utility and ASR remain low. When $N=4$, both utility and ASR reach their optimal states.
Subsequently, as $N$ increases, both metrics decline. When a single synthesized image contains an excessive number of fragmented, spliced patches, model utility borders on collapse, rendering the attack ineffective.This phenomenon occurs because, although a single patch possesses a highly authentic score when $N=1$, its information capacity is limited.
A single image lacks sufficient pixel space and diversity to simultaneously accommodate the global context of the original task and the high-frequency semantics of the hijacking task, thereby constraining both utility and ASR.
When $N=4$, multi-patch splicing introduces data diversity. High-scoring regions from different original samples are condensed into a single image, maximizing its feature density. During gradient computation, the victim model can simultaneously capture sufficiently rich and widely distributed benign and malicious features from these four regions, perfectly supporting the optimization of the trajectory matching loss.
As $N$ continues to increase, the image becomes excessively fragmented (e.g., at a $224 \times 224$ resolution, $N=25$ means each patch contains only $44 \times 44$ pixels).

This excessive fragmentation destroys the structural coherence of the image.
In the OD attack, $N$ determines both the information capacity and the degree of spatial structural fragmentation in a single synthesized image.
The aforementioned experiments thus elucidate the rationale behind setting $N=4$ in our study.

\subsubsection{Impact of Feature Extractor}
Furthermore, we investigated the performance of the OD attack when constructing DODs and executing model hijacking attacks using different feature extractors.
Specifically, we employed ResNet-18 and MnasNet \cite{Tan2018MnasNetPN} as feature extractors.
ResNet-18 was selected to evaluate performance when the feature extractor coincides with the victim model, whereas MnasNet was chosen because it is an evolved version of MobileNetV2 and enables substantial architectural diversity across different network layers, thereby facilitating an evaluation of the OD attack’s generalization and robustness.

As presented in \autoref{tab:feature_extractors}, the experimental results demonstrate that our approach achieves performance comparable to that of MobileNetV2, regardless of dataset resolution (low or high) and irrespective of whether the feature extractor coincides with the victim model.
Moreover, these findings indicate that the OD attack can utilize different models as feature extractors, with only a negligible impact on the model hijacking attack results,  
which remains well within an acceptable tolerance.
Ultimately, this confirms that the OD attack maintains strong generalization capabilities and robustness across different feature extractors.

\begin{table*}[!t]
\centering
\caption{Experiment results using different feature extractors.}
\label{tab:feature_extractors}
\begin{tabular}{c|c|c|c|c|c}
\toprule
\textbf{Original task} & \textbf{Hijacking task} & \textbf{Feature Extractor} & \textbf{Victim Model} & \textbf{Utility $\uparrow$} & \textbf{ASR $\uparrow$} \\
\midrule
\multirow{4}{*}{CIFAR-10} & \multirow{4}{*}{MNIST} & \multirow{2}{*}{ResNet-18} & ResNet-18 & 68.12\% & 98.73\% \\
\cline{4-6}
 & & & VGG16 & 64.48\% & 95.60\% \\
\cline{3-6}
 & & \multirow{2}{*}{MnasNet} & ResNet-18 & 65.25\% & 96.87\% \\
\cline{4-6}
 & & & VGG16 & 63.60\% & 95.52\% \\
\hline
\multirow{4}{*}{ImageNet-Subset (0-99)} & \multirow{4}{*}{ImageNet-Subset (100-199)} & \multirow{2}{*}{ResNet-18} & ResNet-18 & 56.73\% & 69.57\% \\
\cline{4-6}
 & & & VGG16 & 50.17\% & 60.22\% \\
\cline{3-6}
 & & \multirow{2}{*}{MnasNet} & ResNet-18 & 52.13\% & 64.01\% \\
\cline{4-6}
 & & & VGG16 & 49.51\% & 61.80\% \\
\bottomrule
\end{tabular}
\end{table*}

\subsubsection{Impact of Data Dilution}
To further verify the effectiveness and robustness of the OD attack in transfer learning scenarios where distilled datasets are used for fine-tuning, we simulate a common fine-tuning strategy employed by victims who mixing real datasets with distilled datasets.
In our experiments, we vary the proportion of real data used, ranging from $1\%$ to $90\%$.
For the DOD, we configure the IPC to 50.
As shown in \autoref{fig:Mix ratio}, model utility and ASR exhibit similar trends across all 10-class and 100-class scenarios. 
When the proportion of real data ranges from $1\%$ to $50\%$, the increase in utility is negligible; a significant improvement is observed only when the proportion exceeds $70\%$.
Conversely, the ASR experiences only a minor decline within the $1\%$ to $50\%$ range, dropping sharply only when the proportion of real data surpasses $70\%$.

We attribute this phenomenon to the highly condensed nature of dataset distillation.
The DOD generated by the OD attack already encapsulates the vast majority of gradient information required for the original task.
Consequently, introducing additional benign data is redundant for learning the original task; its primary effect is the dilution of malicious features, which leads to a reduction in the ASR.
To achieve a substantial improvement in original task performance, the proportion of real data must exceed $70\%$.
However, this contradicts the typical fine-tuning strategy where victims mix distilled and benign datasets; in practice, victims generally prioritize the distilled dataset, adding only a small fraction of real data during training.
Therefore, the OD attack remains capable of effectively executing the hijacking task even when a significant amount of original data is introduced, demonstrating its strong robustness.

\subsection{Cross-Architecture Transferability}
Based on our defined threat model and the characteristics of real-world transfer learning within open-source dataset supply chains, adversaries typically lack prior knowledge of the model architecture employed by the victim.
In our distillation stage we employ trajectory matching, which relies on the gradient update trajectories of the surrogate model.
Therefore, to verify whether the OD attack remains effective when the victim adopts a model architecture different from that of the surrogate model, we use ResNet-18 as the surrogate model to obtain the training trajectories and employ DenseNet-121 \cite{8099726}, MobileNetV3 \cite{9008835}, and ConvNeXt \cite{liu2022convnet} as victim models to evaluate the cross-architecture transferability of the OD attack.

We adopt CIFAR-10 as the original task and MNIST as the hijacking task.
The experimental results are presented in \autoref{tab:cross_architecture_transfer}.
Regardless of whether the victim model shares the same architecture as the surrogate, the OD attack demonstrates exceptional performance on both Utility and ASR.
This confirms the attack's robust cross-architecture transferability.
The OD attack achieves this by jointly optimizing visual and semantic losses, thereby globally embedding the hijacking task into the deep semantics of the original task.
Victims typically employ models with powerful feature extraction capabilities to achieve high performance on the original task.
Consequently, when fine-tuning such models, they inevitably capture the hijacking semantics, thereby enabling the cross-architecture transferability of the attack.

\begin{table}[!t]
\centering
\caption{Cross-Architecture Transferability of OD Attack.}
\label{tab:cross_architecture_transfer}
\begin{tabular}{c|c|c|c}
\toprule
\textbf{Surrogate Model} & \textbf{Victim Model} & \textbf{Utility $\uparrow$} & \textbf{ASR $\uparrow$}\\
\midrule
\multirow{4}{*}{ResNet-18} & ResNet-18 & 67.84\% & 98.81\%\\
 & DenseNet-121 & 65.73\% &97.90\% \\
 & MobileNetV3 & 63.87\% & 95.50\% \\
 & ConvNeXt-T & 65.38\% & 97.13\% \\
\bottomrule
\end{tabular}
\end{table}

\subsection{OD vs. STRIP}
Although dedicated defense mechanisms against model hijacking attacks are currently absent, we employed the entropy-based backdoor defense mechanism STRIP \cite{10.1145/3359789.3359790} to evaluate the robustness of the OD attack.
STRIP introduces strong perturbations to input samples and utilizes entropy to quantify the randomness of the predicted class distribution.
The underlying assumption is that low entropy indicates the presence of a backdoor, whereas high entropy corresponds to clean inputs.

\begin{figure}[!t]
    \centering
    \includegraphics[width=\linewidth]{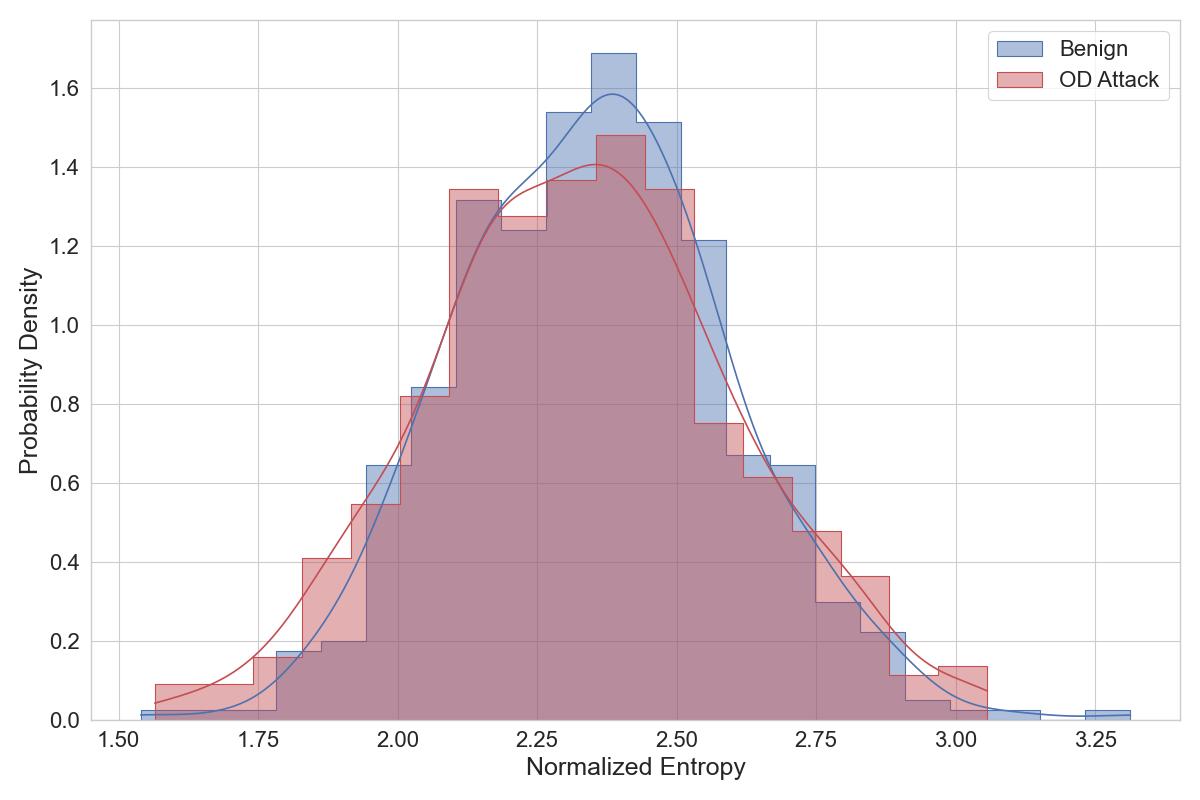}
    \caption{The distribution of entropy of OD attack and entropy of benign samples.}
    \label{fig:strip_od_CIFAR-10_MNIST}
\end{figure}

Based on this assumption, we conducted experiments using CIFAR-10 as the original dataset and MNIST as the hijacking dataset, with an $IPC=50$.
As illustrated in \autoref{fig:strip_od_CIFAR-10_MNIST}, the entropy distribution of the OD attack closely aligns with that of benign samples, exhibiting a high degree of overlap.
This demonstrates that the OD attack effectively circumvents entropy-based detection measures.

The OD attack leverages high-frequency semantic osmosis, a feature that is visually imperceptible yet fragile within the pixel space.
Consequently, superimposing a benign distilled image onto a DOD sample disrupts the high-frequency structures, thereby increasing entropy and enabling evasion of detection.
This confirms both the stealthiness and robustness of the OD attack.
However, it is crucial to distinguish that the OD attack is not a backdoor attack; our methodology does not incorporate any triggers, nor can the embedded adversary-specified task be characterized as such.
Therefore, the development of effective defense mechanisms against model hijacking attacks are an urgent priority.

\subsection{OD vs. DPSGD}
\begin{figure}[!t]
    \centering
    \includegraphics[width=\linewidth]{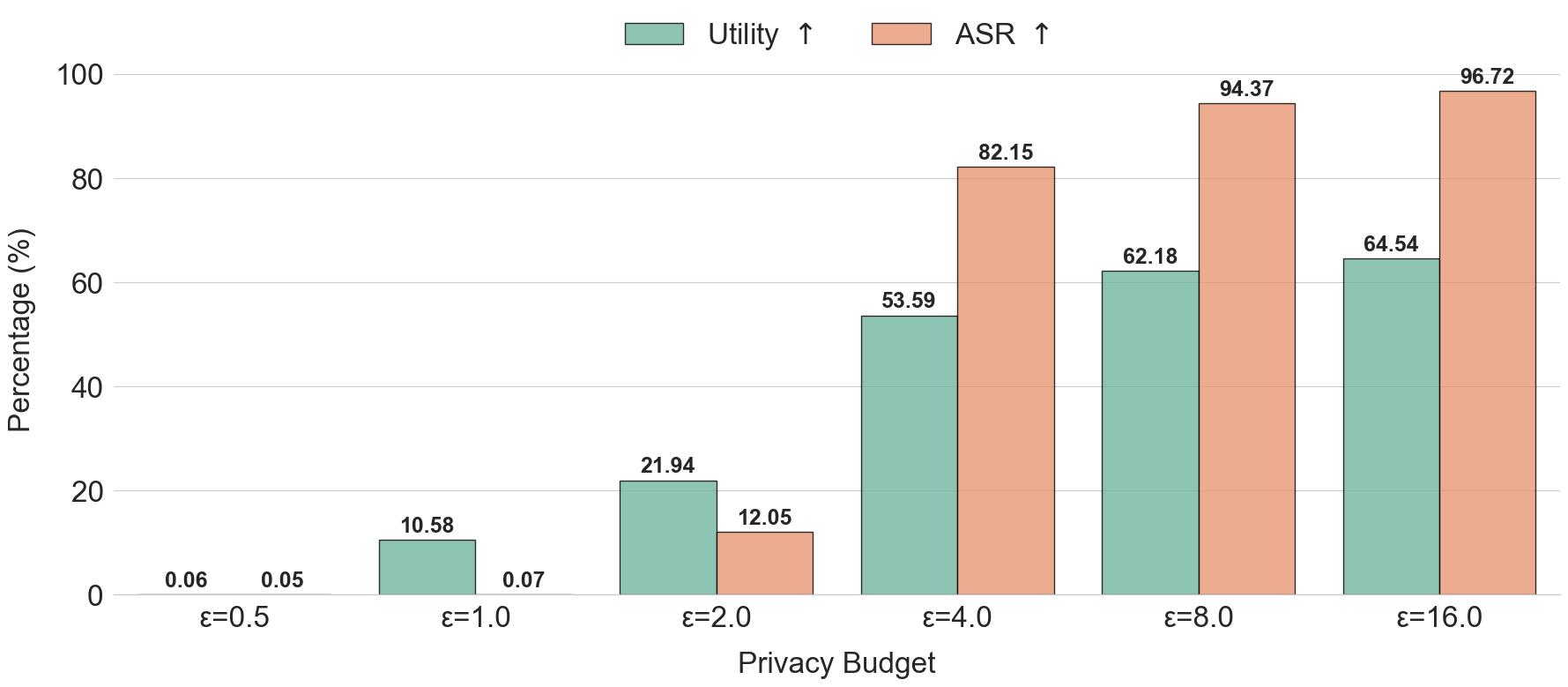}
    \caption{OD under different privacy budgets. ResNet-18 as victim model.}
    \label{fig:DPSGD}
\end{figure}
To further validate the robustness of the OD attack and investigate the efficacy of existing defense mechanisms against it, we fine-tuned ResNet-18 using the Differentially Private Stochastic Gradient Descent (DPSGD) \cite{10.1145/2976749.2978318} defense mechanism across various privacy budgets ($\epsilon \in \{0.5, 1.0, 2.0, 4.0, 8.0, 16.0\}$).
\autoref{fig:DPSGD} presents our experimental results.
Because DPSGD achieves its defensive capabilities via gradient clipping and noise injection, both the utility and ASR experience a precipitous decline under stringent privacy budget constraints.
When the privacy budget is relaxed to 4.0, both utility and ASR begin to recover; at a budget of 8.0, performance approaches the baseline level observed without any defense mechanisms.

These findings indicate that under extremely strict privacy budget conditions, ASR is drastically reduced; however, utility also declines correspondingly, with both metrics approaching zero or random-guessing levels.
Although the ASR is effectively mitigated in this scenario, the severe degradation in utility renders the model ineffective for its original task, making such stringent privacy budgets impractical for real-world deployment.
Conversely, more permissive privacy budgets allow both ASR and utility to recover, demonstrating that our attack successfully withstands the DPSGD defense.

The synchronous changes in ASR and utility arise because the noise introduced by DPSGD destroys the visual feature extraction for the original task and simultaneously disrupts the trajectories of the hijacking task under extremely strict privacy budgets.
Under permissive privacy constraints, the injected noise is reduced, causing the model to inevitably fit the hijacked semantics while learning the original task.
Furthermore, DPSGD cannot selectively distinguish between benign and malicious features. Therefore, utility and ASR necessarily exhibit strong coupling and synchronized scaling effects at specific privacy budget thresholds.

\section{Discussion}
In realistic scenarios, victims could employ third-party distilled datasets from open-source platforms (e.g., Hugging Face \footnote{\url{https://huggingface.co/datasets/devrim/dmd_cifar10_edm_distillation_dataset/tree/main}}, Kaggle  \footnote{\url{https://www.kaggle.com/datasets/ericdeuber/nhl-2nd-period-and-final-scores}}) or purchased from external providers to train or fine-tune models.
However, these victims are typically unaware that such datasets could contain embedded malicious tasks.
This issue is especially pronounced following an OD attack, as the distilled datasets preserve the visual characteristics of the original task, making the presence of hijacking much harder to detect.
Consequently, while distilled datasets can lower training costs, they also introduce risks of model hijacking and, more critically, potential legal liabilities.

\section{Conclusion}
In this paper, we introduce OD attack, a novel model hijacking attack method. 
OD attack integrates model hijacking with dataset distillation, leveraging distilled osmosis samples to significantly reduce the requirement of poisoned samples. 
We evaluate the OD attack across multiple datasets and model architectures.
The experimental results demonstrate that OD attack can successfully execute the hijacking task while minimizing the impact on the performance of the original task. 
We hope that the OD attack serves as a cautionary example to emphasize the security risks that model hijacking poses to dataset distillation and urge caution in using third-party or unverified synthetic datasets.

\bibliographystyle{IEEEtran}
\bibliography{ref}

\appendix
\section{Appendix}

\begin{figure*}[!t]
    \centering
    \begin{subfigure}[b]{0.42\textwidth}
        \centering
        \includegraphics[width=\linewidth]{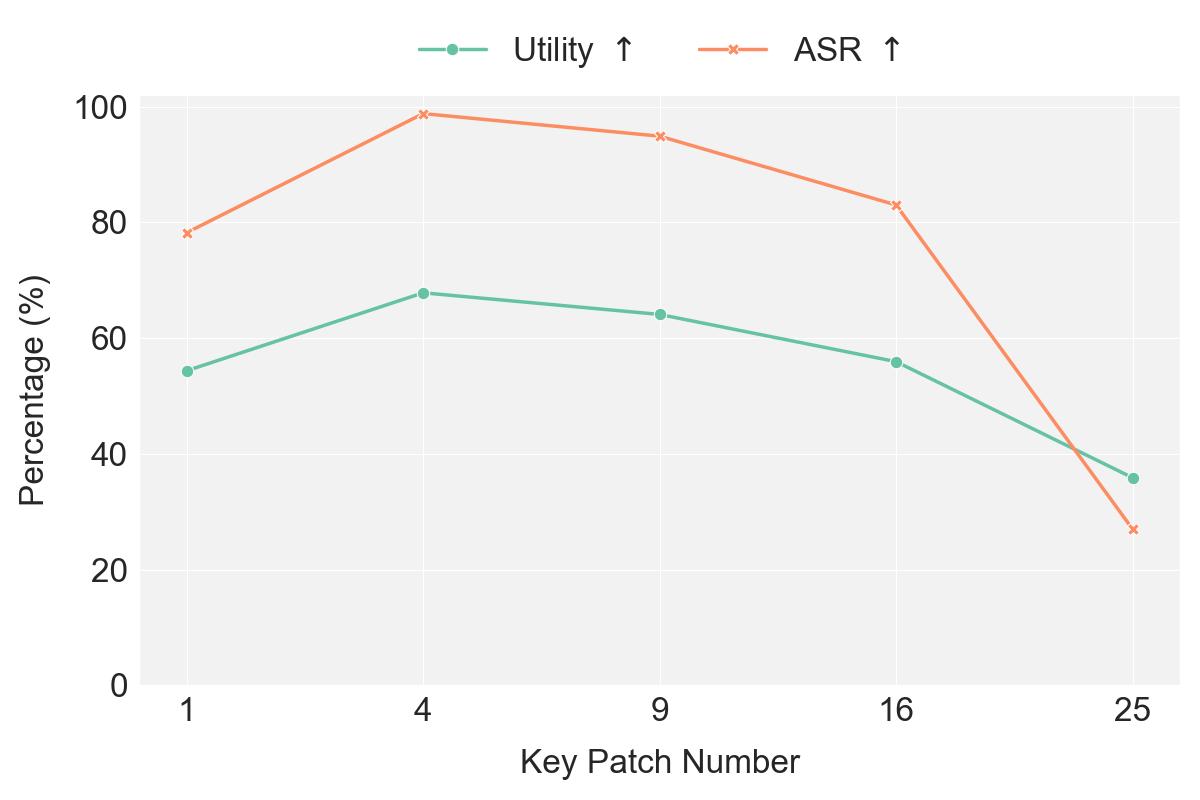}
        \caption{ResNet18: CIFAR-10, MNIST}
        \label{fig:patch_cifar10_mnist_res18}
    \end{subfigure}
    \hspace{8pt}
    \begin{subfigure}[b]{0.42\textwidth}
        \centering
        \includegraphics[width=\linewidth]{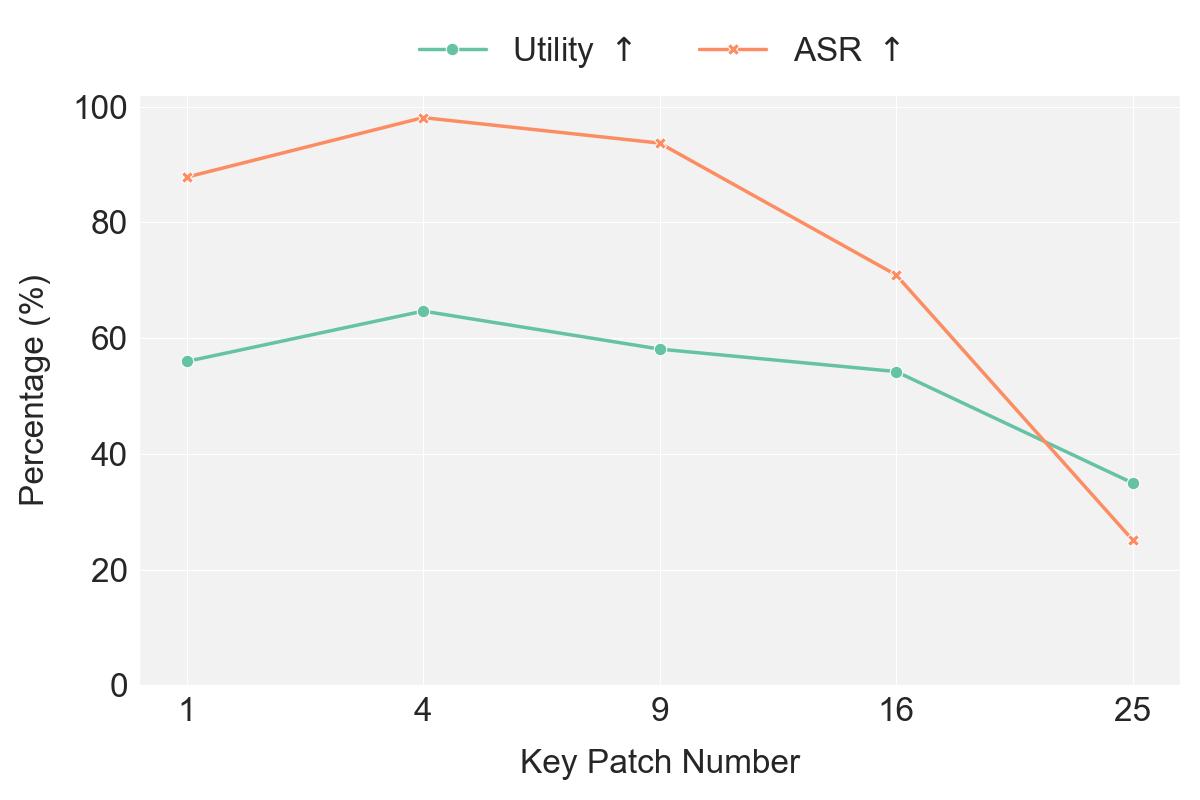}
        \caption{VGG16: CIFAR-10, MNIST}
        \label{fig:patch_cifar10_mnist_vgg}
    \end{subfigure}
    
    \par\medskip 
    \begin{subfigure}[b]{0.42\textwidth}
        \centering
        \includegraphics[width=\linewidth]{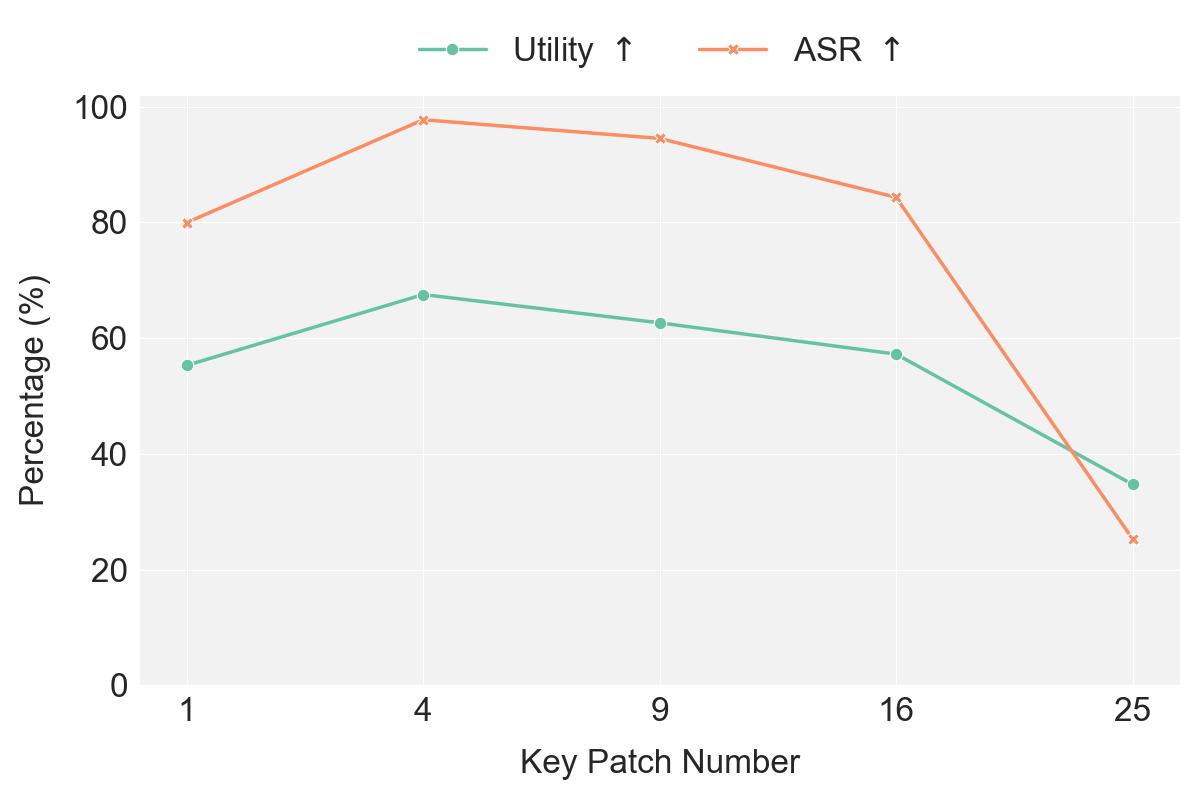}
        \caption{ResNet18: CIFAR-10, SVHN}
        \label{fig:patch_cifar10_svhn_res18}
    \end{subfigure}
    \hspace{8pt}
    \begin{subfigure}[b]{0.42\textwidth}
        \centering
        \includegraphics[width=\linewidth]{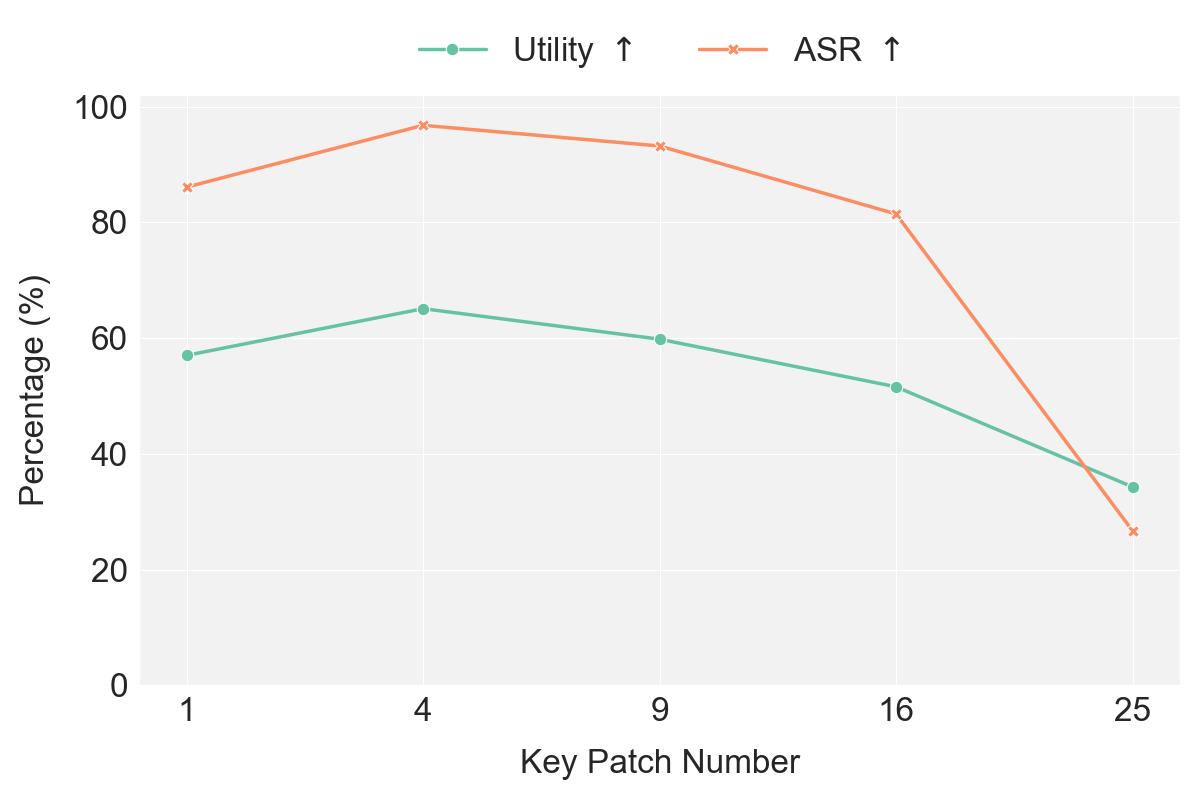}
        \caption{VGG16: CIFAR-10, SVHN}
        \label{fig:patch_cifar10_svhn_vgg}
    \end{subfigure}
    
    \par\medskip 
    \begin{subfigure}[b]{0.42\textwidth}
        \centering
        \includegraphics[width=\linewidth]{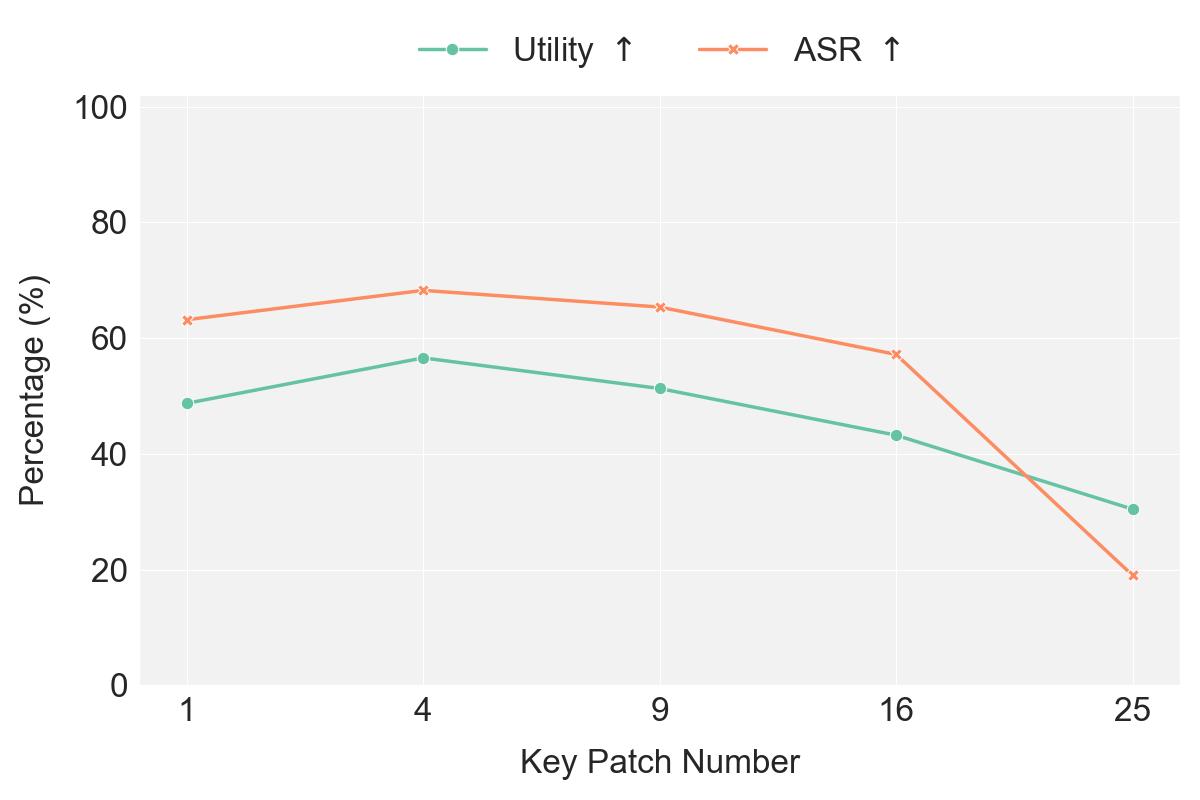}
        \caption{ResNet18: CIFAR-100, Tiny-ImageNet}
        \label{fig:patch_cifar100_tiny_res18}
    \end{subfigure}
    \hspace{8pt}
    \begin{subfigure}[b]{0.42\textwidth}
        \centering
        \includegraphics[width=\linewidth]{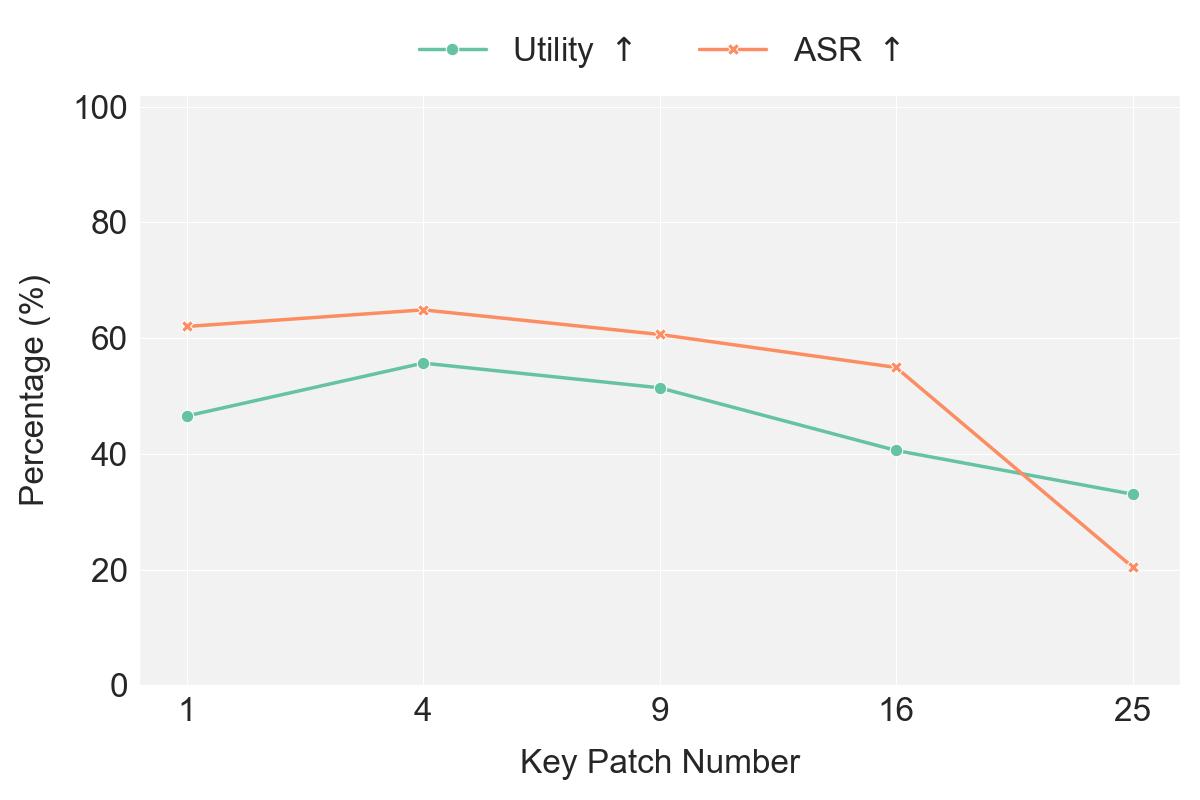}
        \caption{VGG16: CIFAR-100, Tiny-ImageNet}
        \label{fig:patch_cifar100_tiny_vgg}
    \end{subfigure}

    \par\medskip
    \begin{subfigure}[b]{0.42\textwidth}
        \centering
        \includegraphics[width=\linewidth]{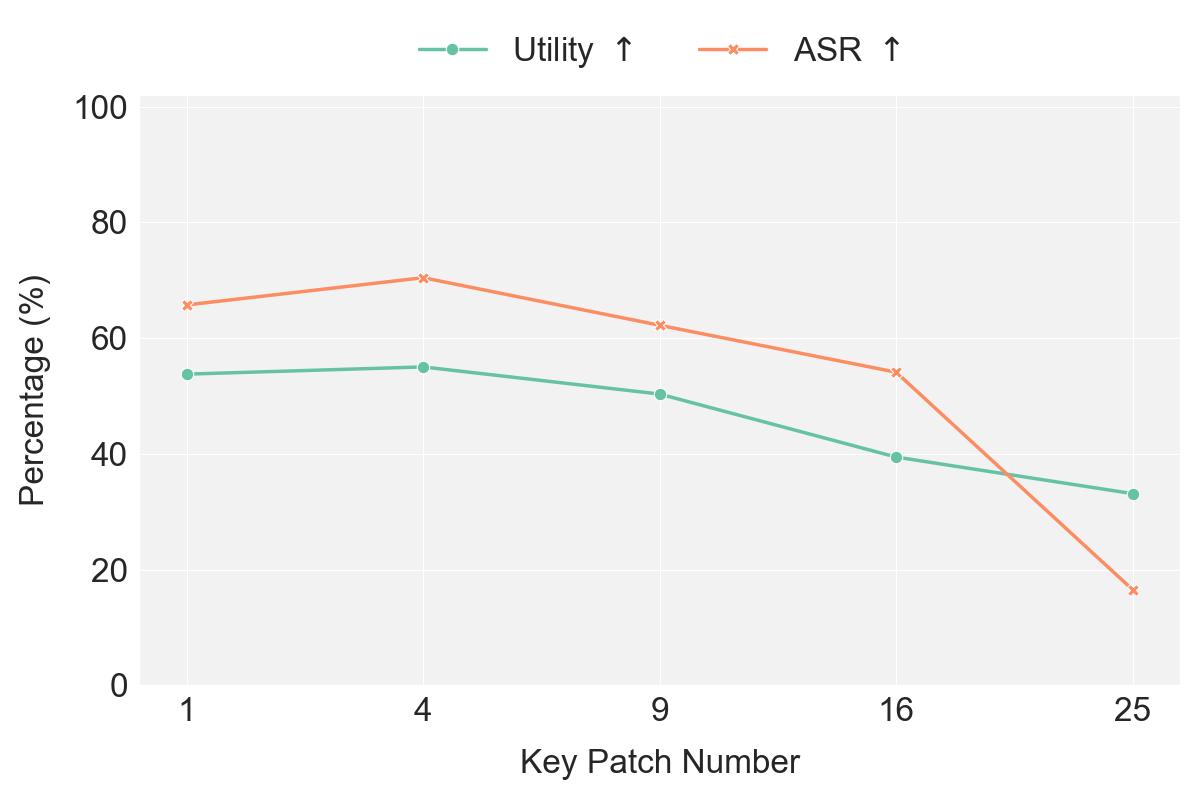}
        \caption{ResNet18: ImageNet}
        \label{fig:patch_image_image_res18}
    \end{subfigure}
    \hspace{8pt}
    \begin{subfigure}[b]{0.42\textwidth}
        \centering
        \includegraphics[width=\linewidth]{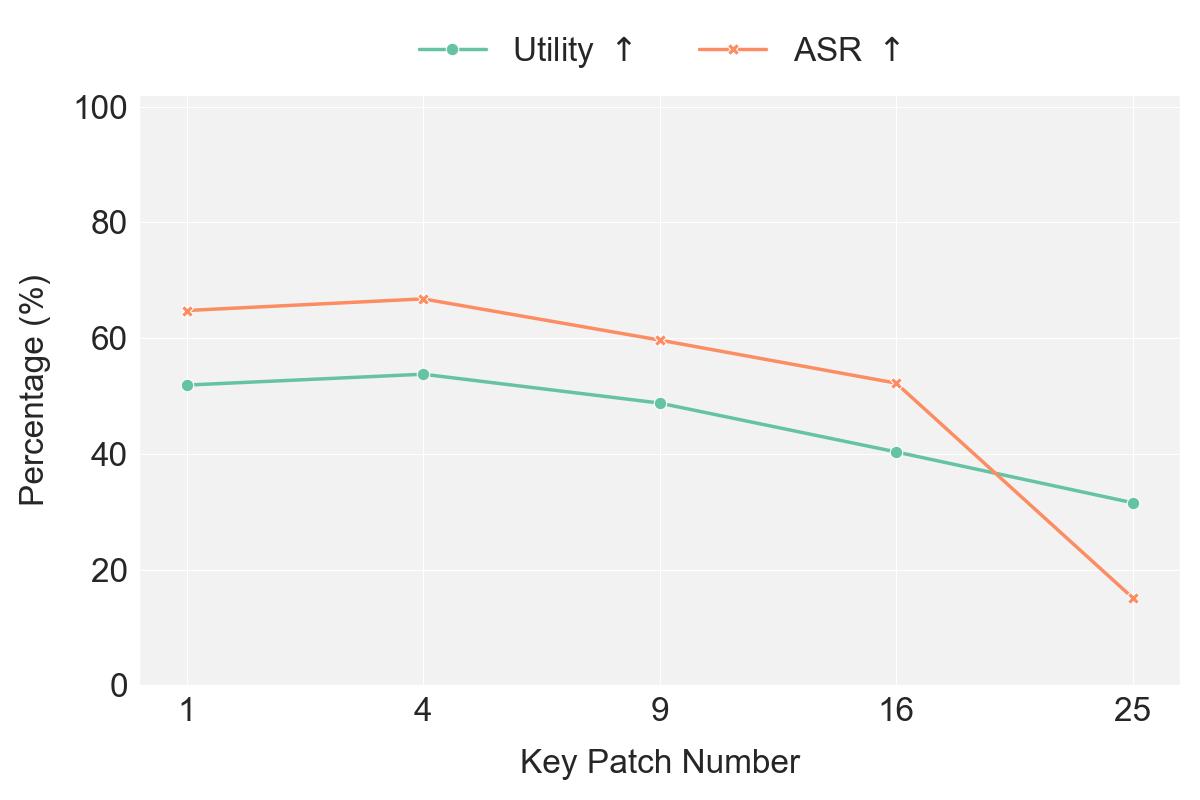}
        \caption{VGG16: ImageNet}
        \label{fig:patch_image_image_vgg}
    \end{subfigure}
    \caption{Experimental results on different key patch numbers.}
    \label{fig:Key Patch}
\end{figure*}

\end{document}